\pdfoutput=1

\documentclass[aps,superscriptaddress,showkeys,nofootinbib]{revtex4-1}
\usepackage{amssymb,amsmath,amsfonts}
\usepackage{graphics}
\usepackage{graphicx}
\usepackage{epstopdf}
\usepackage[pdftex]{hyperref}
\usepackage{longtable}
\usepackage{bm}
\usepackage{subfigure}
\usepackage{color}
\usepackage{setspace}
\usepackage{multirow}

\begin{document}

\title{Global multi-layer network of human mobility}

\author{ Alexander Belyi }
\affiliation{SENSEable City Laboratory, SMART Centre, Singapore}
\affiliation{Faculty of Applied Mathematics and Computer Science,\\ Belarusian State University, Minsk, Belarus}
\affiliation{SENSEable City Laboratory, Massachusetts Institute of Technology, Cambridge, MA, USA}
\author{ Iva Bojic }
 \affiliation{SENSEable City Laboratory, SMART Centre, Singapore}
 \affiliation{SENSEable City Laboratory, Massachusetts Institute of Technology, Cambridge, MA, USA}
\author{Stanislav Sobolevsky \footnote{To whom correspondence should be addressed: sobolevsky@nyu.edu} }
 \affiliation{Center for Urban Science + Progress, New York University, Brooklyn, NY, USA}
 \affiliation{SENSEable City Laboratory, Massachusetts Institute of Technology, Cambridge, MA, USA}
\author{ Izabela Sitko }
\author{ Bartosz Hawelka }
 \affiliation{University of Salzburg, Salzburg, Austria}
\author{ Lada Rudikova }
 \affiliation{Yanka Kupala State University of Grodno, Grodno, Belarus}
\author{ Alexander Kurbatski }
 \affiliation{Faculty of Applied Mathematics and Computer Science,\\ Belarusian State University, Minsk, Belarus}
\author{ Carlo Ratti }
 \affiliation{SENSEable City Laboratory, Massachusetts Institute of Technology, Cambridge, MA, USA}

\begin{abstract}
Recent availability of geo-localized data capturing individual human activity together with the statistical data on international migration opened up unprecedented opportunities for a study on global mobility. In this paper we consider it from the perspective of a multi-layer complex network, built using a combination of three datasets: Twitter, Flickr and official migration data. Those datasets provide different but equally important insights on the global mobility: while the first two highlight short-term visits of people from one country to another, the last one~--- migration~--- shows the long-term mobility perspective, when people relocate for good. And the main purpose of the paper is to emphasize importance of this multi-layer approach capturing both aspects of human mobility at the same time. From one hand we show that although the general properties of different layers of the global mobility network are similar, there are important quantitative differences between them. From the other hand we demonstrate that applications of a multi-layered network can sometimes infer patters which can not be seen from studying each network layer separately. So we start from a comparative study of the network layers, comparing short- and long- term mobility through the statistical properties of the corresponding networks, such as the parameters of their degree centrality distributions or parameters of the corresponding gravity model being fit to the network. We also focus on the differences in country ranking by their short- and long-term attractiveness, discussing the most noticeable outliers. Finally, we apply this multi-layered human mobility network to infer the structure of the global society through a community detection approach and demonstrate that consideration of mobility from a multi-layer perspective can reveal important global spatial patterns in a way more consistent with other available relevant sources of international connections, in comparison to the spatial structure inferred from each network layer taken separately.
\end{abstract}


\maketitle

\section{Introduction}
\label{sec:introduction}

People travel from one country to another for different reasons and while doing so, a lot of them leave their digital traces in various kinds of digital services. This opens tremendous research opportunities through the corresponding datasets, many of which have been already utilized for research purposes, including mobile phone records\,\cite{ratti2006mlu,calabrese2006real,girardin2008digital,quercia2010rse}, vehicle GPS traces\,\cite{santi2013taxi,kang2013exploring}, smart cards usage\,\cite{bagchi2005,lathia2012}, social media posts\,\cite{java2007we,szell2013,frank2013happiness} and bank card transactions\,\cite{sobolevsky2014mining,sobolevsky2014money,sobolevsky2015cities,sobolevsky2015predicting}. By analysing these traces we can reconstruct people movements and afterwards analyse them to see if interesting or useful patterns emerge or to build models for predicting where they will go next.

It has already been shown that results of such analysis can be applied to a wide range of policy and decision-making challenges, such as for example regional delineation\,\cite{ratti2010redrawing,sobolevsky2013delineating} or land use classification\,\cite{pei2014new,grauwin2014towards}. A number of studies focus specifically on studying human mobility at urban\,\cite{gonzalez2008uih,kung2014exploring,hoteit2014estimating}, country\,\cite{amini2014impact} or global scale. When considering aspects of human mobility at global scale, in particular two major types of movements can be observed: an international migration\,\cite{greenwood1985human,fagiolo2013international,abel2014quantifying,tranos2012international} and short-term trips explored through for example geo-localized data from Twitter\,\cite{hawelka2014} or Flickr\,\cite{paldino2015flickr,sobolevsky2015scaling}.

Some studies tried to primarily explain and model global mobility\,\cite{greenwood1985human,fagiolo2013international,abel2014quantifying,tranos2012international}, while other rather focused on its applications, such as revealing the structure of the global society through global mobility networks\,\cite{ratti2010redrawing,sobolevsky2013delineating,hawelka2014}. Some scholars even considered relationships between human migration and economical links between countries\,\cite{sgrignoli2015relation,fagiolo2014does}. However, global human dynamics by itself has a complex nature containing various types of mobility, including different processes as permanent relocation and short-term visits; and thus cannot be fully understood through any single data source focusing on just one particular aspect of human behavior. On the other hand, recent studies provided methodological background to deal with multi-layer complex networks\,\cite{kivela2014multilayer}.

In this study we use three different sets of data each representing different kind of people movements. Namely, Flickr and Twitter represent short-term human mobility, while migration network contains information about a long-term one. Although Flickr and Twitter are similar in a certain sense, they show different types of people activity during travel motivated by different reasons: in case of Flickr it is mostly activity during a leisure time and visits to some touristic places, while in case of Twitter it can be just any kind of activity: business or touristic one. 
Moreover, they are complement as in some counties where only one of those services may be popular and widely used by people. 

Further we provide a comparative study of different layers of human mobility. The specific focus of our study is on demonstrating that such a complex approach to human mobility considering it from different short- and long-term perspectives is of a principle importance: a multi-layer global mobility network shows patterns not seen from each layer separately. In order to evaluate our hypothesis, we applied a method for detecting communities in multi-layer networks and compared outcomes with those for the other existing international connections (e.g., language, colony and trade networks). The results showed that communities detected in the three-layer network are on average much more similar to existing international connections than the ones observed in each layer separately.

\section{Datasets}
\label{sec:networks}

As our study aims for investigating human mobility from two different perspectives (i.e., long-term and short-term), we include three datasets where two of them capture short-term human movements such as touristic, personal or business travel, and one of them reflects long-term mobility such as people moving to another country to live there. In that sense, short-term human mobility is revealed from more than $130$~million geo-tagged digital objects (e.g., videos and photographs) publicly shared on Flickr and more than $900$ million geo-tagged tweets posted by $13$ million users on Twitter, while long-term one is inferred from United Nations official migration statistics. Moreover, in order to compare mobility patterns determined in the aforementioned way and in some sense also to verify our results, we used three other datasets showing international connections: colonial dependency, network of languages shared by countries and network of international trade.

Flickr dataset used in our study contains more than $130$ million photographs and videos. It was created by merging two publicly available Flickr datasets~--- one coming from a research project and another from Yahoo\,\cite{flickr1,thomee2015yfcc100m}. The records in two datasets partially overlap, but since each digital object in both datasets has its id, we were able to merge them by omitting duplicates and choosing only those records that were made within a 10 year time window, i.e., from 2005 and until 2014. The second dataset on short-term mobility consists of geo-tagged messages posted during 2012 and collected from the digital microblogging and social media platform Twitter. Data was collected with the Twitter Streaming API\,\cite{twitterapi} and cleansed from potential errors and artificial tweeting noise as previously described in\,\cite{hawelka2014}.

In order to build a two-layer directed and weighted network that describes short-term human mobility, we had to convert Flickr and Twitter datasets into origin-destination matrix where origins represented users' home countries and destinations are places (i.e., countries) where users created digital objects or tweeted from. Since both datasets do not contain information about user home location, we had to determinate which of the users are acting in each location as residents by using the following criteria: a person is considered to be a resident of a certain country if this is the country where he/she took the highest number of the photographs/videos over the longest timespan (calculated as the time between the first and last photograph taken within the country) compared to all other countries for the considered person\footnote{More about why it is important to choose the right home detection method for the given dataset and which methods are the most commonly used ones can be found in~\cite{bojic2015choosing}.}.

Using this simple criteria, we were able to determine home country for over $500$ thousand users in Flickr dataset that took almost $80\%$ of all the photographs/videos in the dataset (i.e., more than $90$ millions in total), while the rest of users for which home country could not be defined mostly belong to a low-activity group taking photographs only occasionally. When constructing the two-layer weighted and directed mobility network, we only considered users for whom we were able to determine their home country. Finally, two countries are connected with a link if there is at least one person from the first country that had some activity in the second country where the value of every weighted link in this network corresponds to the total number of users from one country that made digital objects or tweeted in the other one.

\begin{figure}[t!]
\centering
\subfigure[\label{fig::penetration_flickr}Flickr]{\includegraphics[width=.49\textwidth]{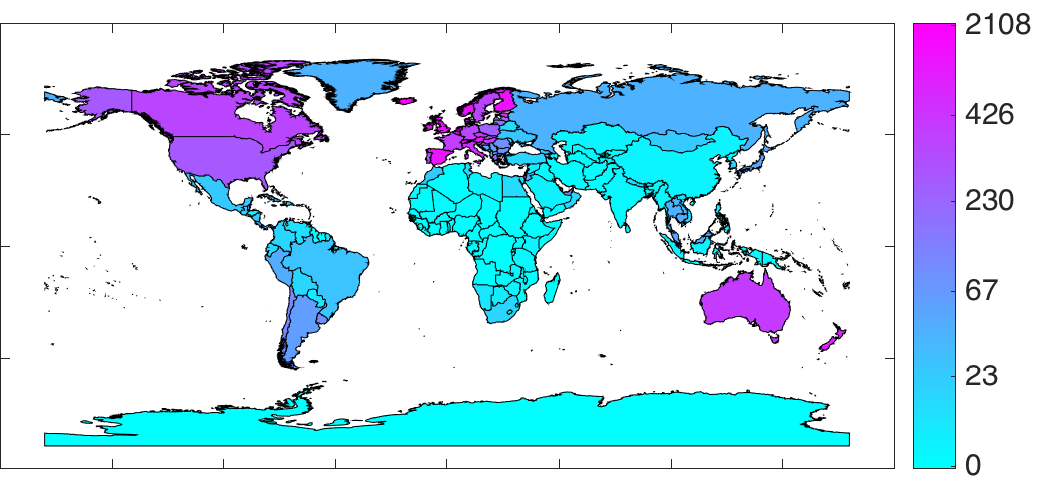}}
\subfigure[\label{fig::penetration_twitter}Twitter]{\includegraphics[width=.49\textwidth]{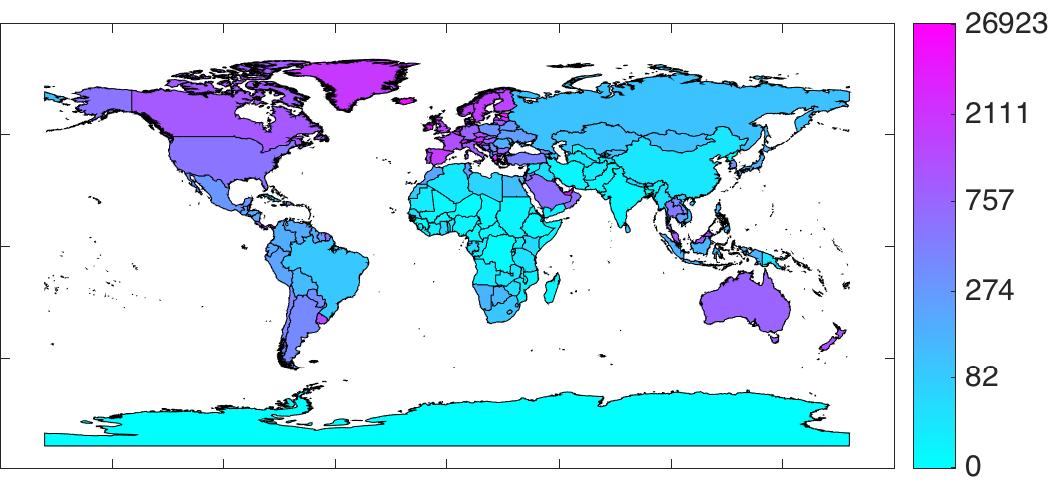}}
\caption{\label{fig::penetration}Penetration of Flickr and Twitter into countries all over the world as number of users who travel abroad per $1$ million of population.}
\end{figure} 

We should mention here that Flickr and Twitter are much more widely used in developed countries while penetration into some other countries can be quite low. Figure\,\ref{fig::penetration} shows how many users per $1$ million of population from each country we determined to be active outside their homeland. We can see that penetration in China (mostly due to restrictive legal regulations) and India as well as in most African counties is pretty low. About $75\%$ of all the countries for Flickr and $45\%$ for Twitter have less than $0.01\%$ of population ever using these resources abroad.

The third layer was constructed using dataset with statistics on number of foreign citizens or foreign-born population living in each country in July 2010. This data is publicly available and can be downloaded from United Nations, Department of Economic and Social Affairs website\,\cite{unMigration}. This statistic is basically already provided in a form of an origin--destination matrix, so the process of extending the two-layer directed and weighted network on human mobility with one additional layer describing long-term human movements was very straightforward.

Finally, to see how the built three-layer mobility network correlates with cultural and economical parameters, we created three separate networks of country relationships based on colonial dependence, common language spoken by people in different countries and bilateral trade between countries. In the first network of colonial dependence\,\cite{hensel2009icow} two countries are connected if one of them was a colony or dependent territory of another one. Furthermore, the second network is network of common spoken languages\,\cite{Languages} where two countries are connected if there is at least one common language that is official in both countries or spoken by majority of population in both of them. Unlike the first two, the third one is a weighted network of trade flows between countries obtained from United Nations Commodity Trade Database~\cite{UNCOMTRADE} where the value of the link represents the amount of import/export (in US dollars) in 2012.

\section{Quantitative and qualitative properties of mobility networks}
\label{sec:analysis}

Before we get to the analysis of the entire three-layer mobility network, constructed as explained in the previous section, we start from a comparative study of some basic quantitative characteristics of the network layers. We try to figure out their differences and similarities, seeing if those different datasets effectively tell us the same stories about global human mobility or the different ones. Specifically, the main focus of this section will be the comparison of long-term versus short-term attractiveness of the countries. Focusing on the countries' ability to attract foreign visitors, for this analysis we excluded loop-edges from all the networks and considered a common measure of `incoming degree centrality', which in our case of weighted networks becomes `strength centrality'. We additionally consider the distribution of links' weights in order to gain insights from the overall composition of the international mobility fluxes.

\begin{figure}[t]
\centering
\subfigure[\label{fig::strDistr}]{\includegraphics[width=.49\textwidth]{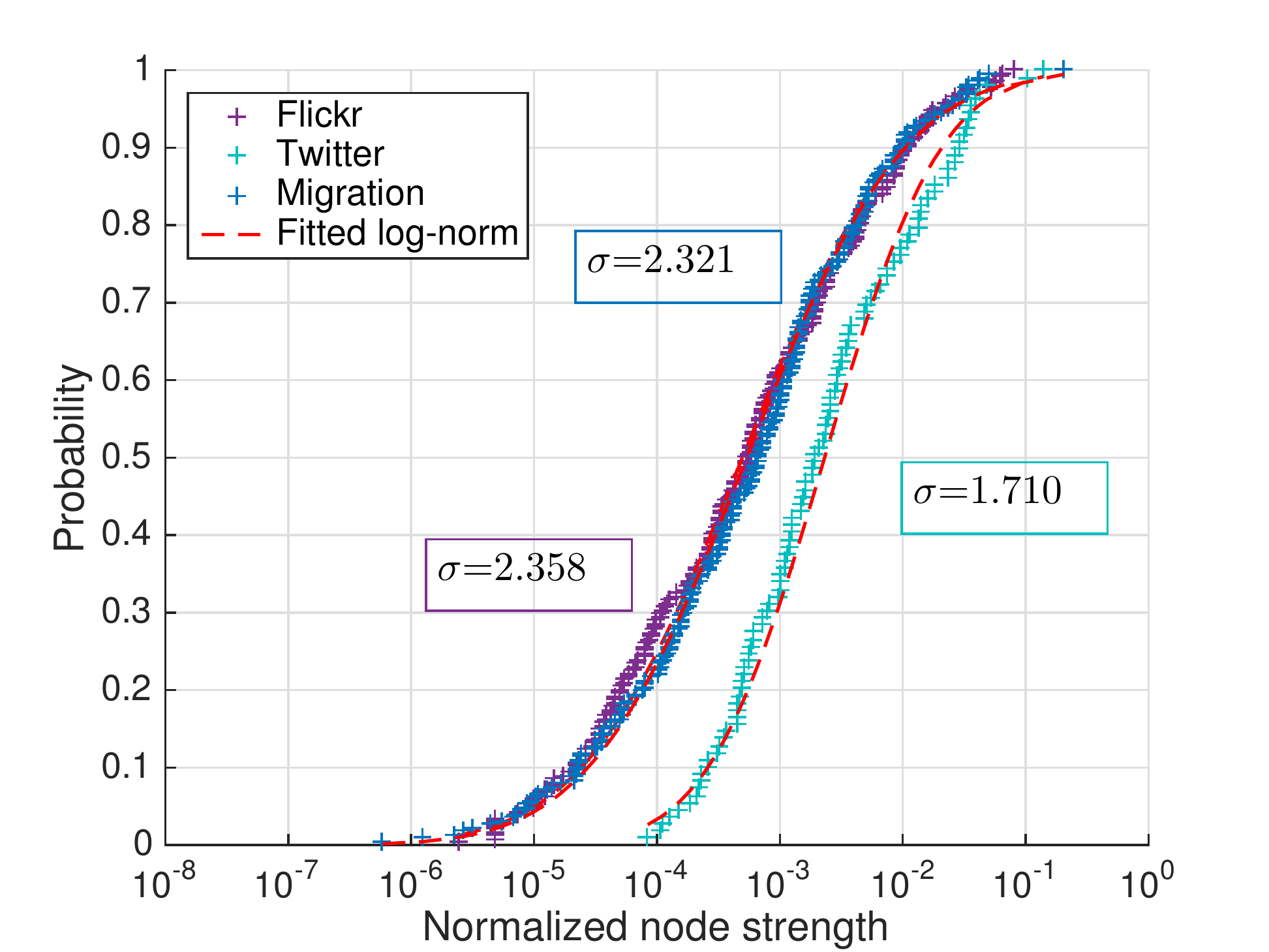}}
\subfigure[\label{fig::weightDistr}]{\includegraphics[width=.49\textwidth]{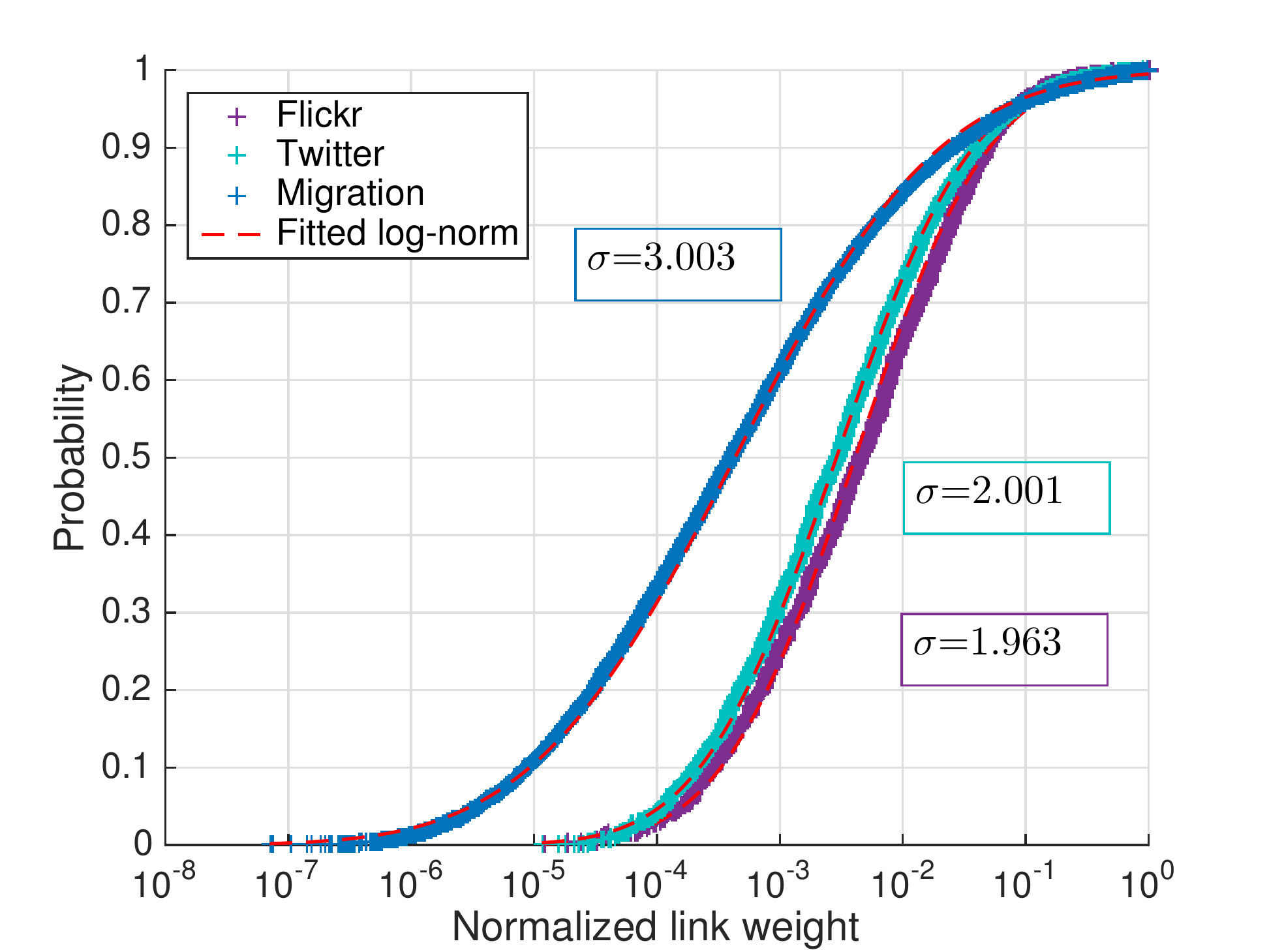}}
\caption{\label{fig::cumulative}Cumulative distribution of normalized node strength~(a) and normalized link weight~(b).}
\end{figure}

\begin{figure}[b]
\centering
\includegraphics[width=0.55\textwidth]{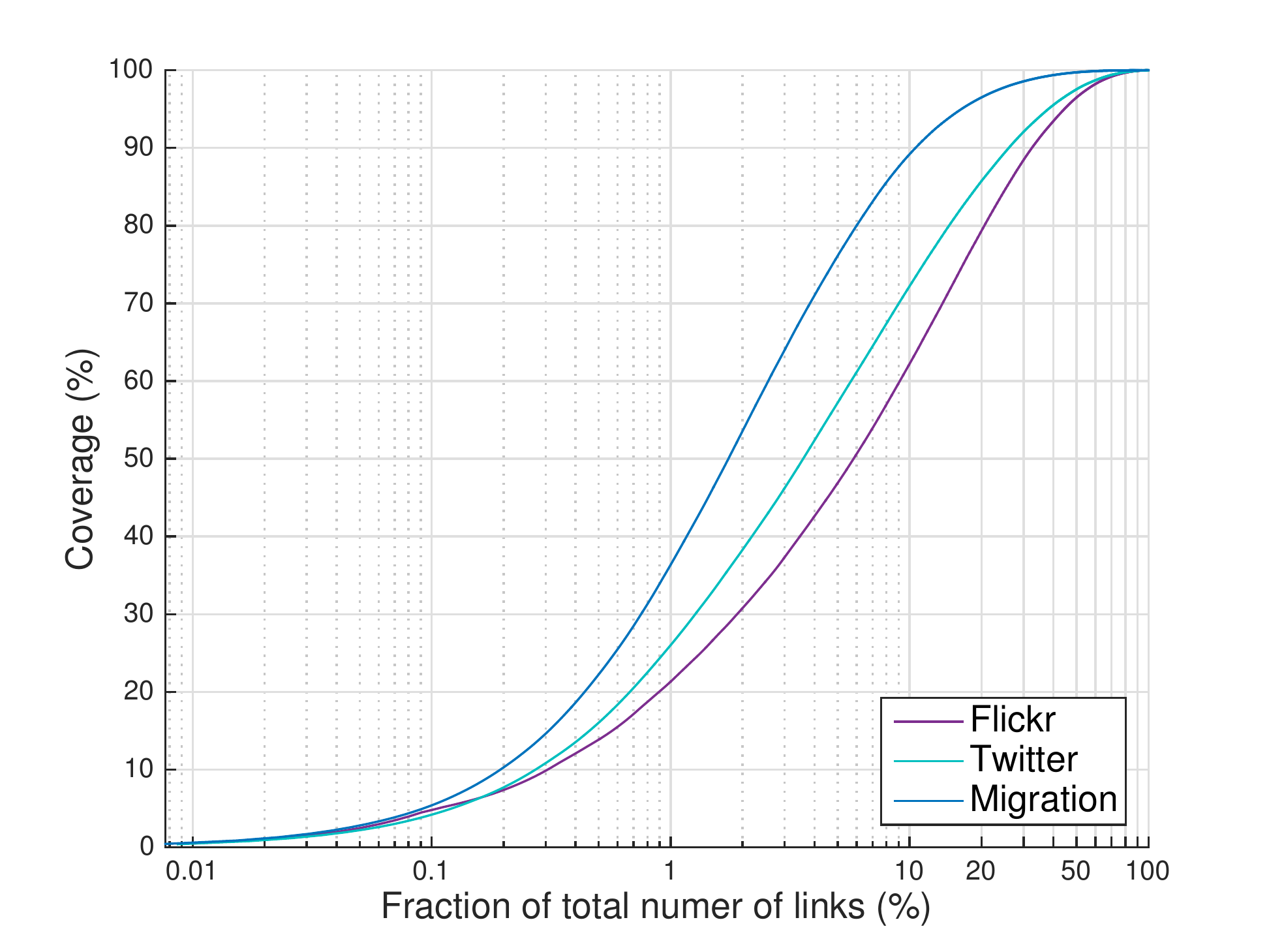}
\caption{\label{fig::flowCoverage} Flow coverage.}
\end{figure}

Relative country attractiveness for the foreigners could be computed as the fraction of all people who travel outside their country of origin which come to that considered country of destination and can be calculated as a normalized node incoming strength (i.e., as the sum of weights of all incoming non-loop edges to the given destination divided by the sum of all non-loop edge weights in the network). We also consider relative weights for each specific mobility flux between the two countries as the number of people moving between them normalized by the total number of people moving out of the considered origin. From Figure\,\ref{fig::cumulative}, which plots cumulative distribution function of normalized incoming strengths of the nodes and relative link weights for all three networks, we can conclude that all shown distributions are pretty similar to log-normal, with the migration network links showing a value of variance much higher compared to other two touristic networks. This means that the migration fluxes are generally more diverse than the short-term mobility ones as seen from Flickr and Twitter networks.

\begin{figure}[b]
\centering
\includegraphics[width=0.9\textwidth]{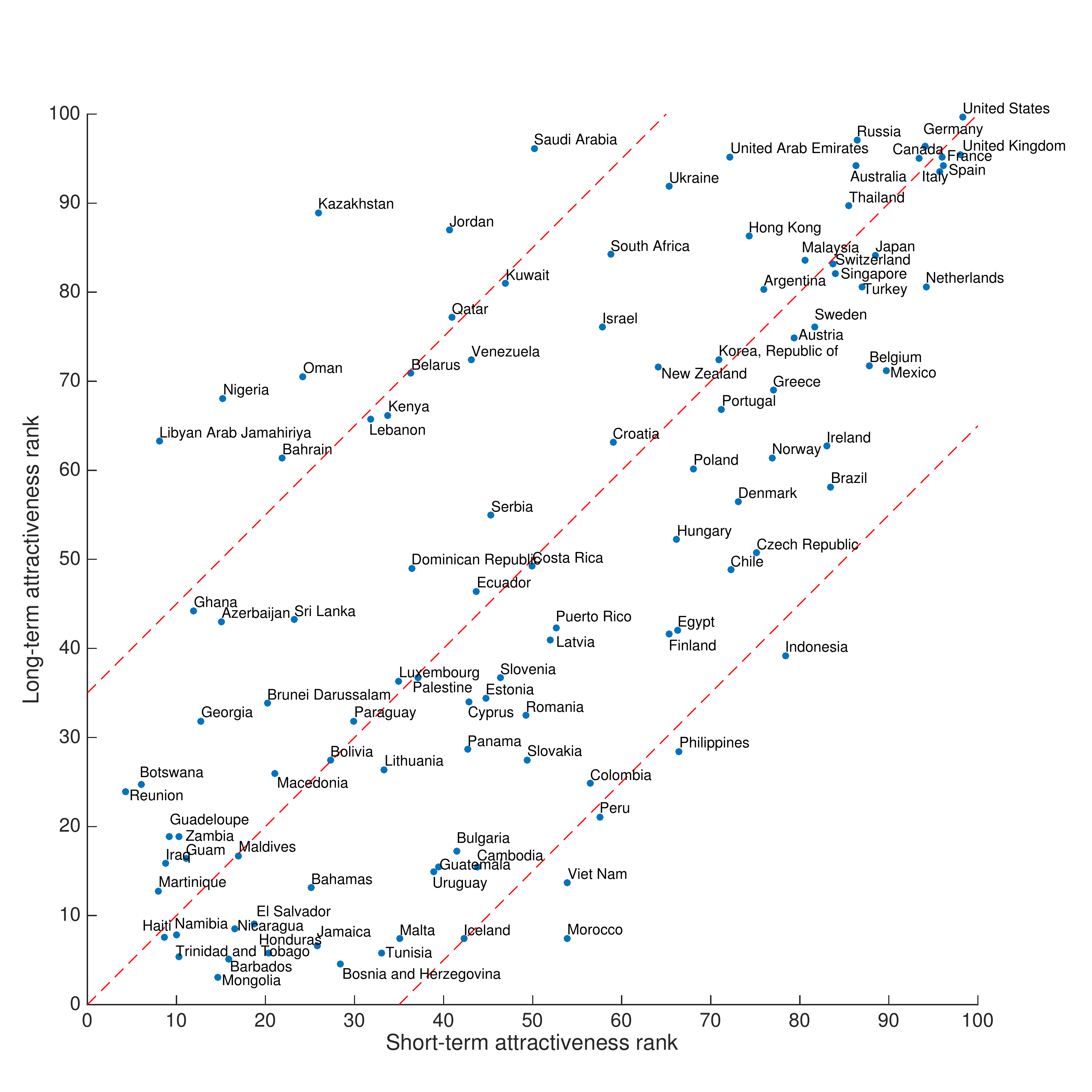}
\caption{\label{fig::ranks}Comparison of counties' short-term vs. long-term attractiveness ranks.}
\end{figure}

In order to explore the observed link diversity in more details, on Figure\,\ref{fig::flowCoverage} we show how many links one needs to cover a certain percentage of the entire network's flow. We can see that in the migration network top $1\%$ of links cover almost $40\%$ of entire network flow, and $10\%$ of links cover $90\%$, while those values are much smaller for other two networks. This observation can be explained by the fact that although migration links are generally diverse, for tourism there is a much broader choice of major destinations than for migration.

Finally, we calculate countries' ranks according to their foreign attractiveness. Figure\,\ref{fig::ranks} shows how a short-term (averaged over Twitter and Flickr networks) attractiveness rank correlates with a migration attractiveness rank. However interesting stories are told by the outliers. Here we can see some very clear patterns: on one side there are prosperous Middle East countries attractive for professional migration, but relatively less popular for short-term visitors, including tourists, and on the other side there are less prosperous countries not so much attractive for migrants, but offering much more interesting natural attractions which makes them primary touristic destinations. Needless to say, there are also highly developed countries with very well established tourism, plenty of business visitors and lots of incoming migrants, being highly ranked in both short and long-term attractiveness.

Although results of our analysis presented in this section showed that all three networks share some common properties, at the same time they differ in many aspects. Common features of networks reflect certain facts about different types of countries: the most developed countries are popular for both types of visitors, but then again there are also countries attractive only from one side. Moreover, it has been shown that the migration flow is much more concentrated around just couple of pairs of origin/destination countries, while tourists have a larger variety of choices when deciding where to go next.

\section{Modeling mobility}
\label{sec:models}

One of the goals of many studies on human mobility has always been to predict mobility flows. Although couple of different models were proposed, still most of related work rely on classical gravity model~\cite{Gravity1946Zipf,Barthelemy2011Spatial}. The model takes spatial population distribution including distances between different locations as input and predicts mobility fluxes with respect to several model parameters, called exponents. The least are either assumed either are to be fit from partial knowledge of the network. Recently an alternative  parameter-free radiation model has been suggested~\cite{RadiationModel}, which allows to predict human mobility just based on the spatial distribution of the country population without any parameters to fit, although the radiation model in turn relies on some partial knowledge of the mobility network as specified below. We compare performance of those models on our global mobility network and use them to reveal and compare patterns behind the three layers of this multi-layer network. 

Classical gravity model tries to predict number of people moving from origin $i$ to destination $j$ as $w_{ij}=C\frac{pop_i pop_j}{d_{ij}^\alpha}$, where $d_{ij}$ is the distance between $i$ and $j$, $C$ is a global normalization constant ensuring that the predicted total activity is the same (or on the same scale) as observed and $\alpha$ is an adjustable parameter of the model. To address the network heterogeneity we used the total amount of outgoing mobility $s_i^{out}$ observed in the network instead of the population $pop_i$ of origin $i$, as the least might not be the most relevant parameter for our networks due to high differences in penetration level of Flickr and Twitter across populations of different countries. This will also help a fair comparison between gravity and radiation models as the least as we will describe below, specifically relies on the knowledge of $s_i^{out}$. Therefore, in our case the final expression for predicting a flux from $i$ to $j$ is 
$$
w_{ij}=C\frac{s_i^{out} pop_j}{d_{ij}^\alpha}.
$$ 

We also consider a `locally-normalized' version of the gravity model, i.e., gravity model in a form 
$$
w_{ij}=\frac{s_i^{out} pop_j d_{ij}^{-\alpha}}{\sum_{k\ne i}pop_k d_{ik}^{-\alpha}}.
$$

This type of constrained model is rooted in the earlier work~\cite{wilson1967statistical} and more recently presented in~\cite{sagarra2013statistical} and~\cite{grauwin2015identifying}.

Unlike gravity, radiation model is claimed to be parameter-free. It uses only population distribution to predict the flux of people as 
$$
w_{ij}=s_i^{out}\frac{pop_i pop_j}{(pop_i + s_{ij})(pop_i + pop_j + s_{ij})},
$$
where $s_{ij}$ is the population within circle with center at $i$ and radius equal to the distance between $i$ and $j$ excluding population of $i$ and $j$, $s_i^{out}$ as before represents the total number of the commuters from $i$. Worth mentioning however that the model still depends on the knowledge of $s_i^{out}$. Masucci~et~al.~\cite{masucci2013gravity} adjusted the radiation model, introducing the appropriate normalization factor for finite systems. After incorporating this factor we came up with equation that we used in our experiments in a form of 
$$
w_{ij}=\frac{s_i^{out}}{1 - \frac{pop_i}{\sum_{k}{pop_k}}}\frac{pop_i pop_j}{(pop_i + s_{ij})(pop_i + pop_j + s_{ij})}.
$$

\begin{figure}[t]
\centering
\subfigure[\label{fig::fit_flickr}Flickr]{\includegraphics[width=.32\textwidth]{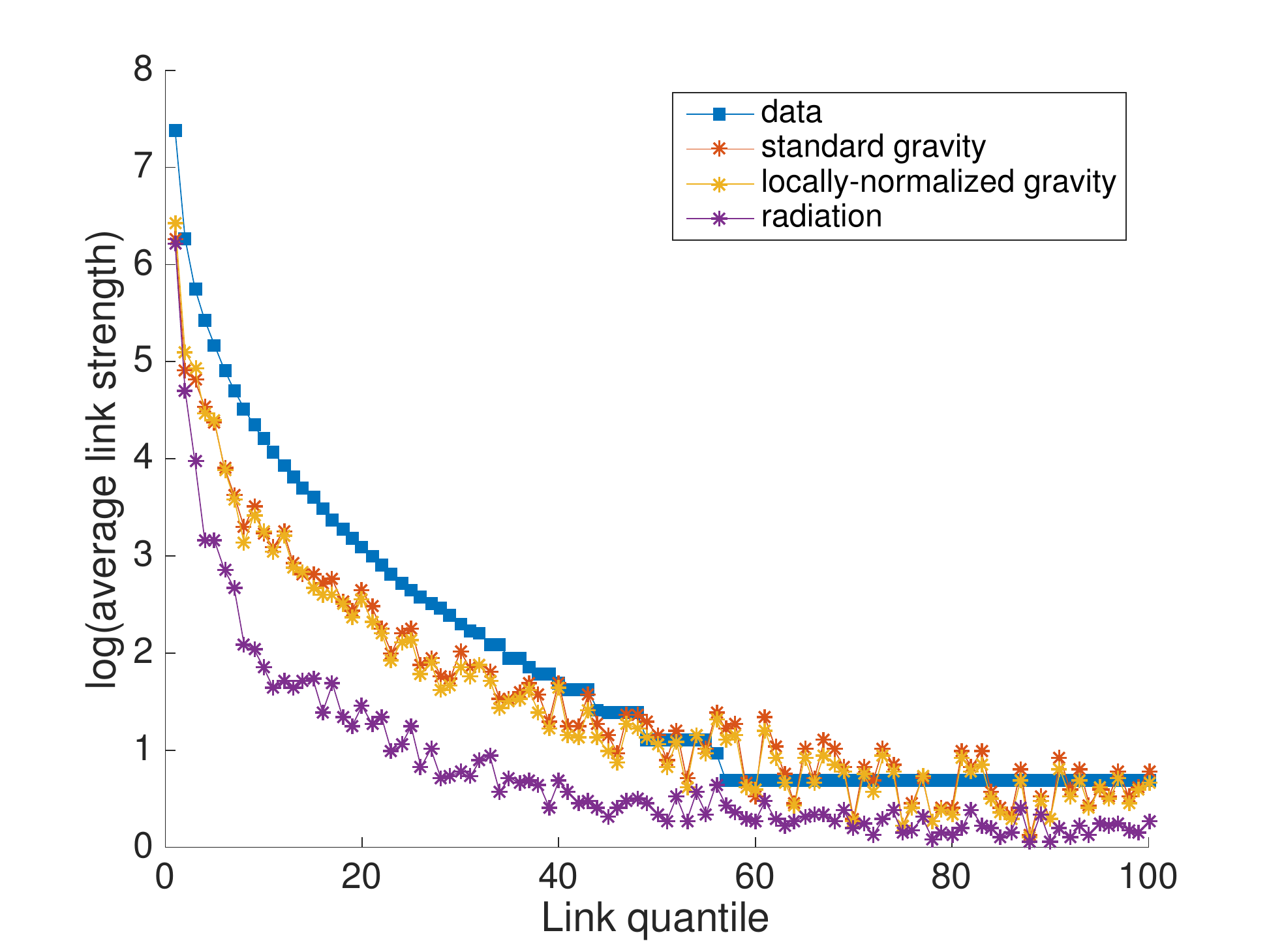}}
\subfigure[\label{fig::fit_twitter}Twitter]{\includegraphics[width=.32\textwidth]{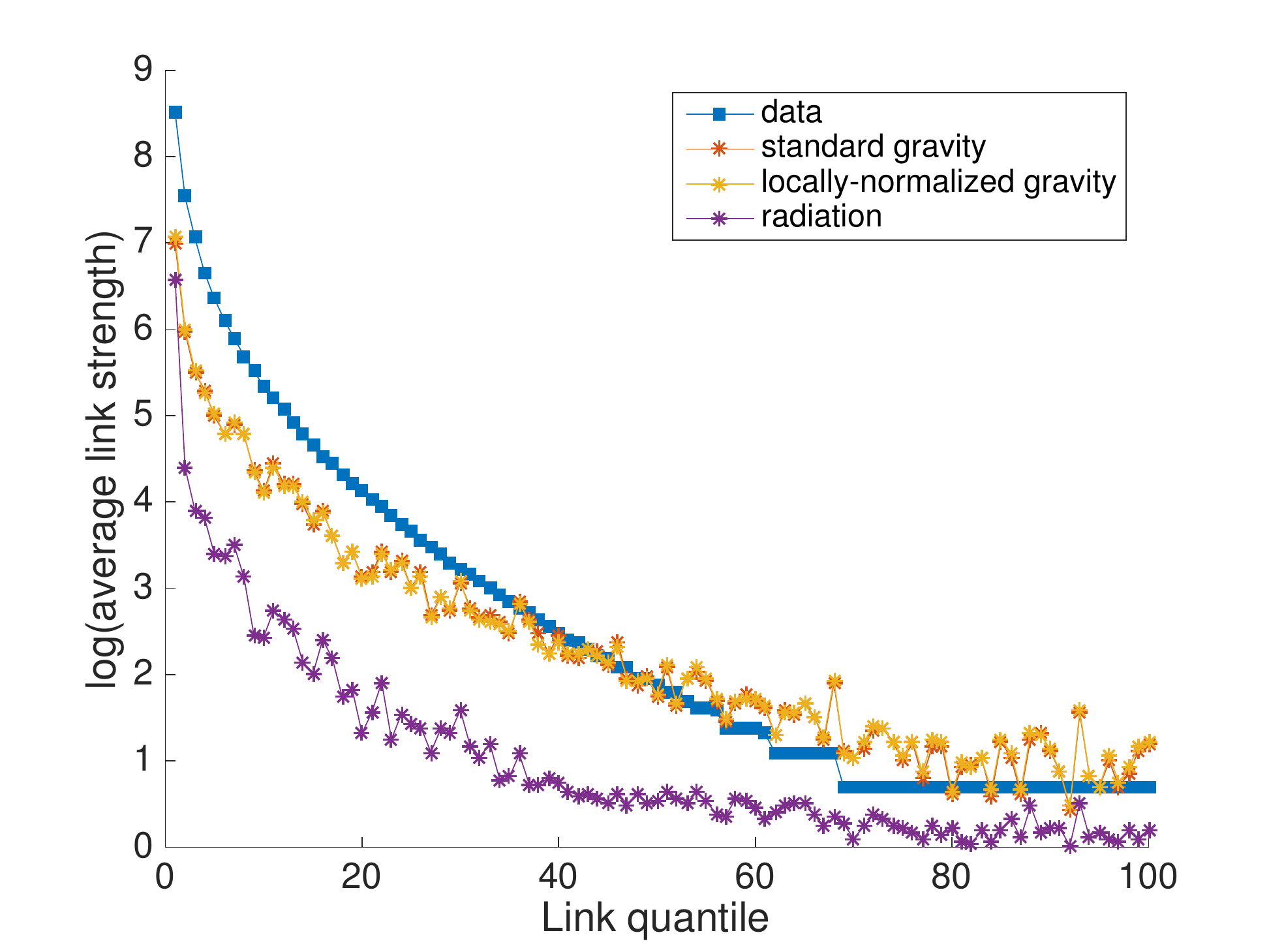}}
\subfigure[\label{fig::fit_migration}Migration]{\includegraphics[width=.32\textwidth]{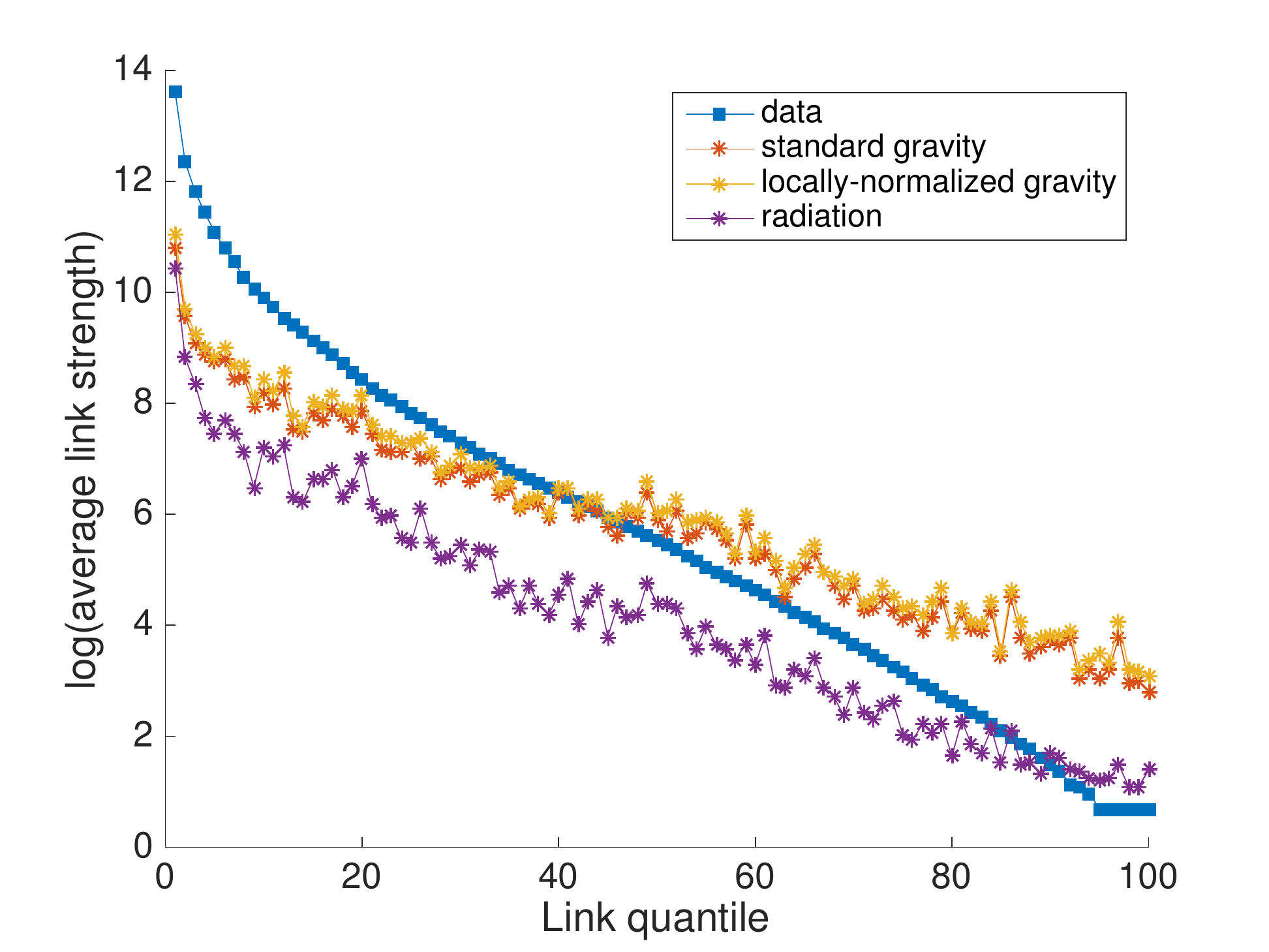}}
\caption{\label{fig::models_fit}Fit of models to Flickr, Twitter and migration networks.}
\end{figure}

After fitting the gravity model and its `locally-normalized' version on the logarithmic scale together with constructing the radiation model for the three layers of our multi-layer mobility network we got results presented in Figure\,\ref{fig::models_fit}. From the figure and values of $R^2$ presented in Table\,\ref{tab::fit_results} we can conclude that for all three layers the gravity model~--- both, classical and locally-normalized version~--- fit much better for the appropriate choice of the parameters.

For~that~reason we will stick to the gravity model in our further analysis.
Worth mentioning that sometimes other parameters are introduced to the gravity model as well, such as population exponents. However when fit those got pretty close to $1$ and did not improve the model performance much, so we limit our analysis with having just the distance exponent making the highest impact and revealing the strongest pattern.

\begin{table}[t]
\caption{\label{tab::fit_results}Results of fitting models to layers.}
\centering
{
\begin{tabular}{|l|c|c|c|c|}
\hline
Parameter                                & Flickr  & Twitter & Migration  \\ \hline
$\alpha$ gravity                         & 1.2934  & 1.1016  & 2.0188     \\ \hline
$\alpha$ `locally-normalized' gravity    & 1.4404  & 1.1153  & 2.1995     \\ \hline
$R^2$ gravity                            & 0.59    & 0.72    & 0.72       \\ \hline
$R^2$ `locally-normalized' gravity       & 0.61    & 0.72    & 0.71       \\ \hline
$R^2$ radiation                          & 0.34    & 0.28    & 0.55       \\ \hline
\end{tabular}
}
\end{table}

And the above pattern happens to be pretty distinctive for different layers of the network as one can see from the Table\,\ref{tab::fit_results}, where we report values obtained while fitting the parameter $\alpha$, i.e., the exponents for the impact of the distance between origin and destination countries. Those indicate how fast the flux of people moving between the two countries decays with increment of distance between them. For both~--- classical and locally-normalized versions of the model~--- the relative pattern looks quite consistent: the exponent is slightly higher for Flickr network compared to Twitter and much higher for the migration layer. As the higher is the exponent, the faster is the decay of mobility with the distance, we can conclude that migrations are usually much more local, while short-term mobility is much less spatially constrained.

\section{Detecting communities in the three-layer mobility network}
\label{sec:community_detection}

After exploring properties of each network (layer) separately, which revealed certain general similarities, but also noticeable differences between the layers, we wanted to check our hypothesis that the three-layer mobility network shows patterns that cannot be observed if looking only layer by layer. In order to evaluate the hypothesis we focused on the problem of community detection. 

Previous studies~\cite{ratti2010redrawing,sobolevsky2013delineating} have shown that community detection in human interaction and mobility networks usually leads to connected spatially cohesive communities (even with no spatial considerations in the community detection method) often revealing meaningful geographical patterns. There was no exception for the global mobility networks estimated from Twitter\,\cite{hawelka2014}, as well as from migration data\,\cite{fagiolo2013international}. However, while separate layers of mobility network provide interesting partial insights on the spatial structure of the global human society, we wonder if certain patterns can be seen only from the multi-layered perspective.
Several ways to detect communities in multi-layer networks were proposed recently\,\cite{mucha2010community,tang2012community} and for our study we chose the approach based on a direct multi-layer generalization of the most widely used objective function for network partitioning which is modularity\,\cite{newman2004,newman2006}.

Even before building a multi-layer generalization of the modularity, one more adjustment has to be made to it for our case in order to account for the absence of loop edges in the mobility networks we consider. For that purpose we altered the way null-model used by modularity estimate weight of each edge. In its classical form modularity uses $\frac{s_i t_j}{\sum_{k}{s_k}}$ as an expected weight of the edge from $i$ to $j$, where if $w_{ij}$ is the weight of the link from $i$ to $j$ then $s_i=\sum_j w_{ij}$ and $t_j=\sum_i w_{ij}$. This can be explained as that distribution of the outgoing weight $s_i$ among all the possible destinations is proportional to their incoming weight $t_j$. However, if loop edges do not participate in this distribution then it should be rather
$\frac{s_i t_j}{\sum_{k\ne i}{s_k}}$ or $\frac{s_i t_j}{\sum_{k\ne j}{t_k}}$, depending on whether it is seen as distribution of the outgoing weight $s_i$ among all the destinations except $i$ itself, or as distribution of the incoming weight $t_j$ among all the origins except $j$. Finally, as a final estimation we use average of those, that leads to the expression
${1 \over 2} \left(\frac{s_i t_j}{m - t_i} + \frac{s_i t_j}{m - s_j}\right)$, where $m = \sum_{k}{s_k} = \sum_{k}{t_k} = \sum_{ij}{w_{ij}}$ is total weight of all edges.

Since it has already been shown that modularity suffers from certain drawbacks, such as a resolution limit\,\cite{Fortunato02012007ResolutionLimit,Good2010PerformanceOfModularity} preventing it from recognizing smaller communities, we also used the approach proposed by Arenas~et~al.~\cite{Arenas2008Analysis} that involves introduction of a so-called resolution parameter, leading to the further adjustment of the modularity score. This way the final formula for the adjusted modularity measure used for our case of the mobility networks free of the loop edges is:
\[
	\frac{1}{2m}
	\sum\limits_{i\neq j}
		\left(
			2w_{ij} - a \frac{s_i t_j}{m - t_i} - a \frac{s_i t_j}{m - s_j}
		\right)
	\delta(C_i,C_j),
\]
where $a$ denotes the resolution parameter, $i,j$ are nodes, $C_i, C_j$~-- the communities they belong to, $\delta(x,y)=1$ if $x=y$, $0$ otherwise.

To deal with the multi-layer network, where all layers share the same nodes, following Tang~et~al.~\cite{tang2012community} we combined adjusted modularity scores of each layer taking their average value and using this as a resulting utility function for the multi-layer network as following:
\[
	Q = {1 \over 3}
		\sum_{l=1}^3{
			{1 \over 2m^l}
			\sum_{i,j}{
				\left( 2w^l_{ij} - a \frac{ s^l_i t^l_j }{ m^l - t^l_i } - a \frac{ s^l_i t^l_j }{ m^l - s^l_j } \right)
			}
			\delta\left( C_i, C_j\right)
		},
\]
where $l$ denotes layer, $w^l_{ij}$ is the weight of the link from $i$ to $j$ in layer $l$, $s^l_i=\sum_j w^l_{ij}$, $t^l_j=\sum_i w^l_{ij}$, $m^l = \sum_{ij} w^l_{ij}$. While in order to find the best partitioning we optimized this multi-layered version of modularity using efficient and precise Combo algorithm\,\cite{Combo}, suitable for dealing with different types of objective functions.

\begin{figure}[b]
\centering
\includegraphics[width=0.65\textwidth]{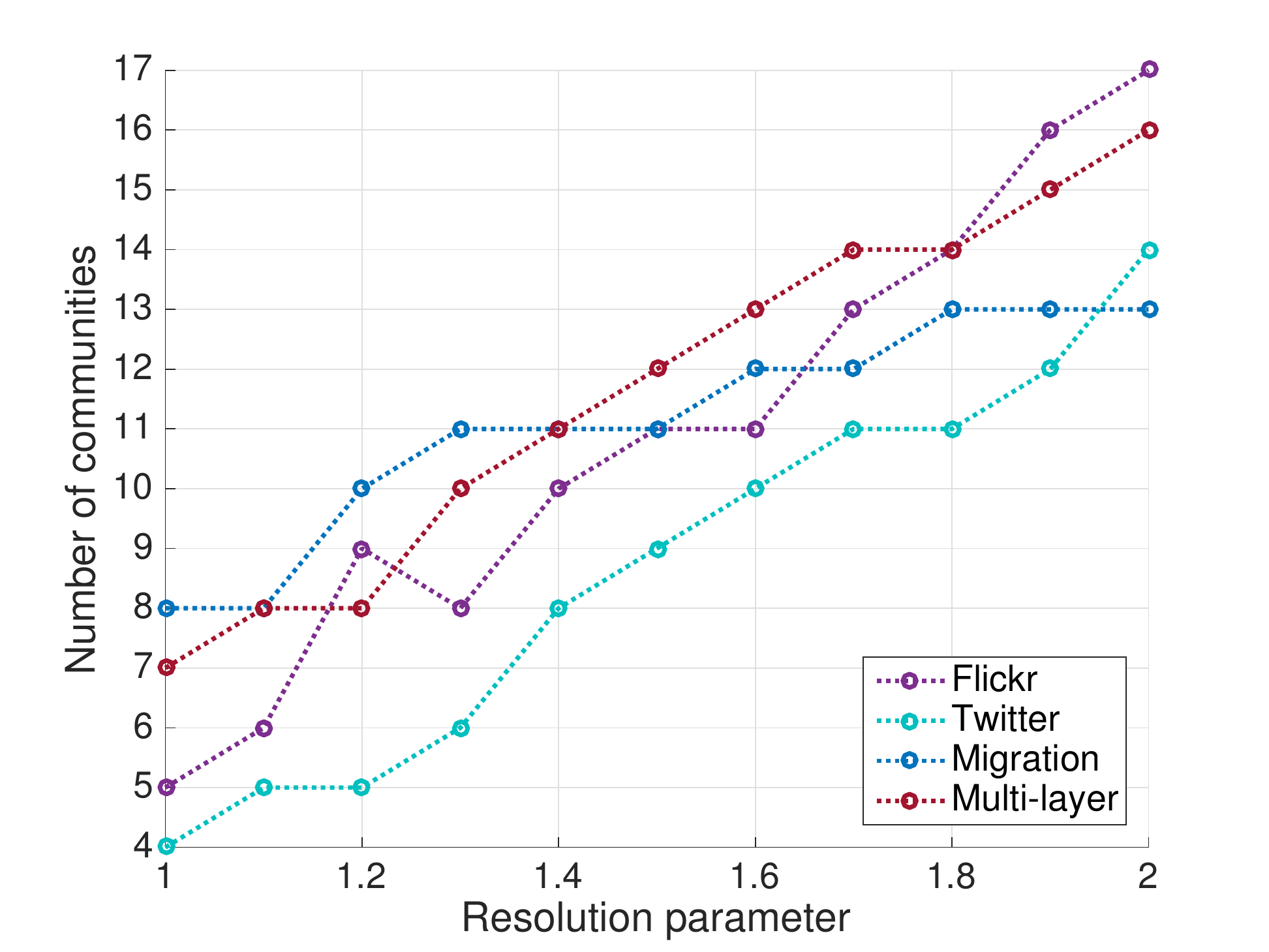}
\caption{\label{fig::comm_number}Number of communities depending on resolution parameter.}
\end{figure}

\begin{figure}[t!]
\centering
\includegraphics[width=.49\textwidth]{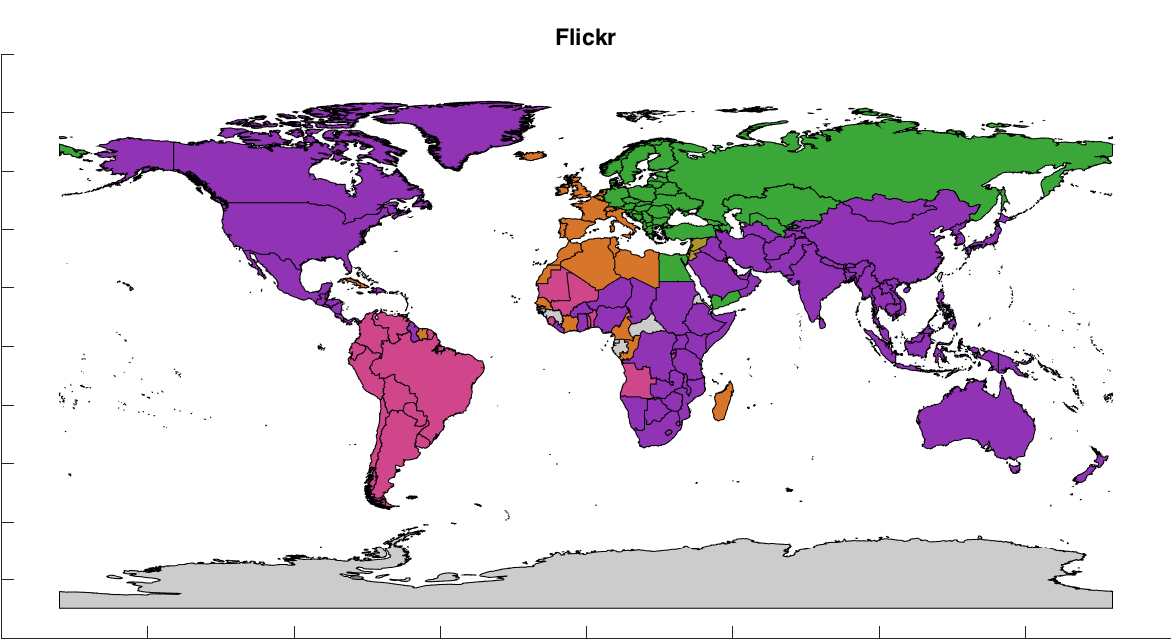}
\includegraphics[width=.49\textwidth]{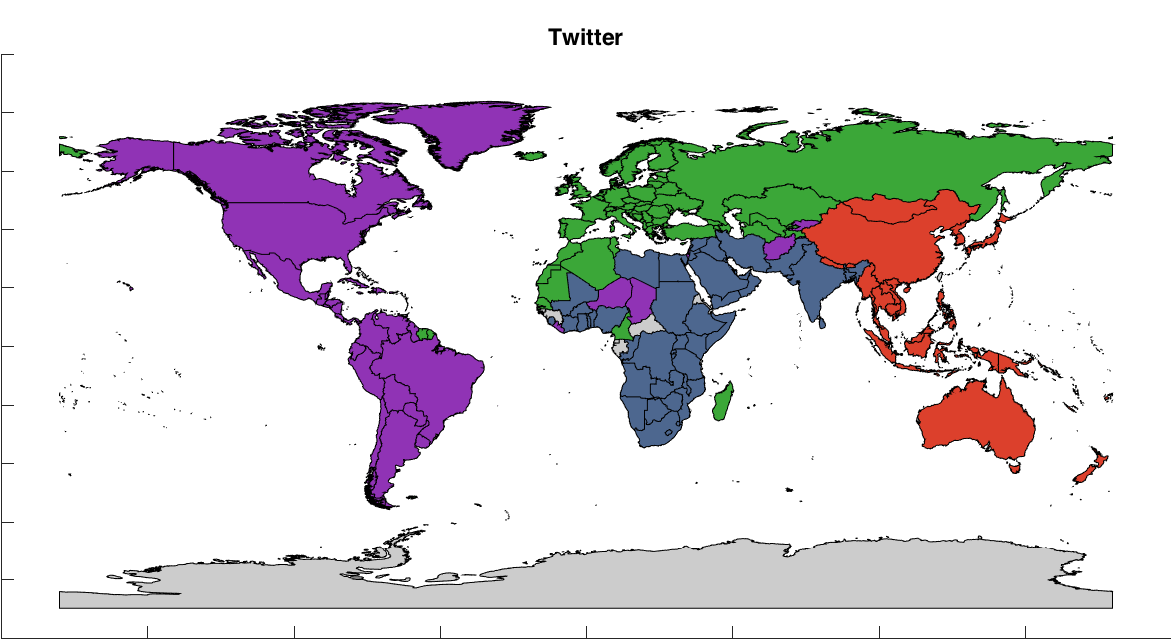}
\includegraphics[width=.49\textwidth]{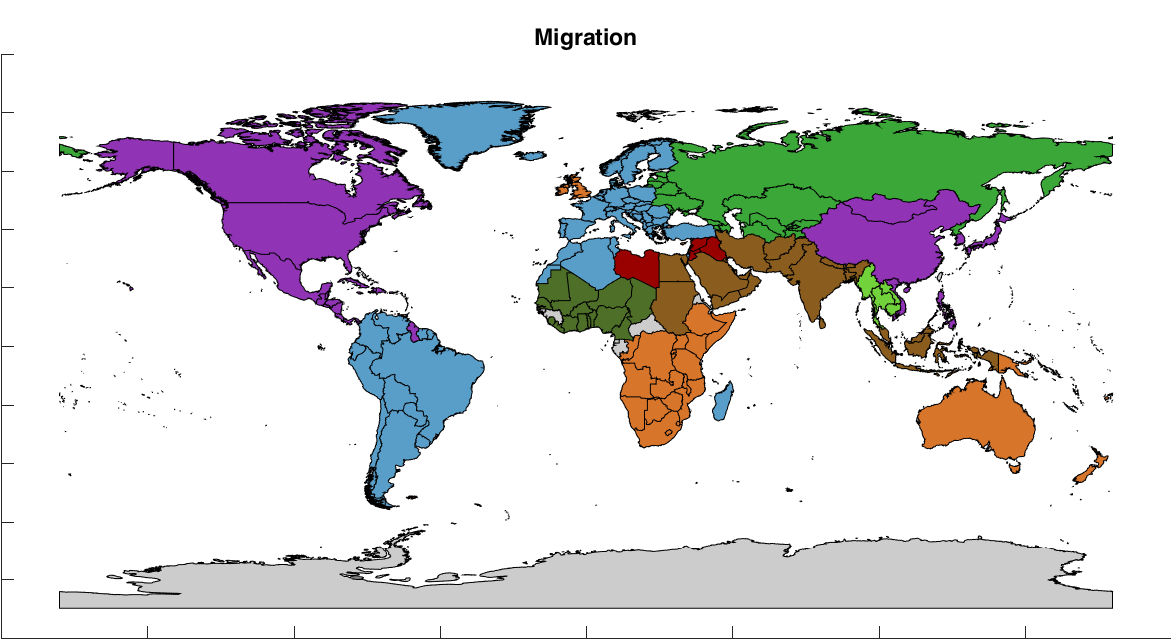}
\includegraphics[width=.49\textwidth]{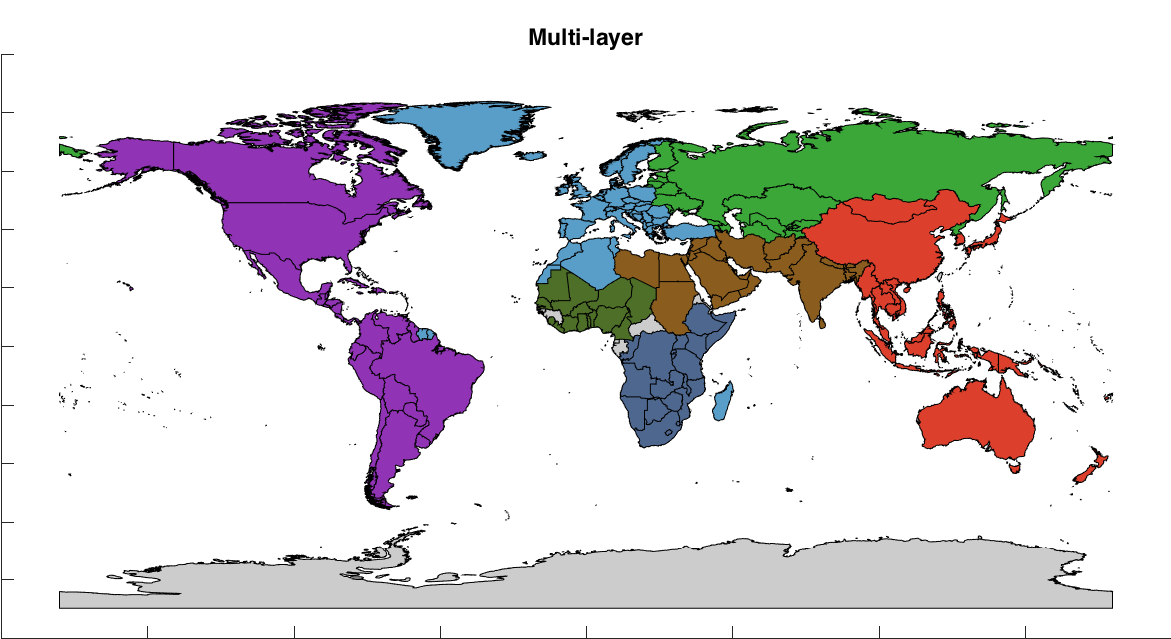}
\caption{\label{fig::partitioning10}Communities for resolution parameter value equal~to~$1.0$.}
\end{figure}

\begin{figure}[b!]
\centering
\includegraphics[width=.49\textwidth]{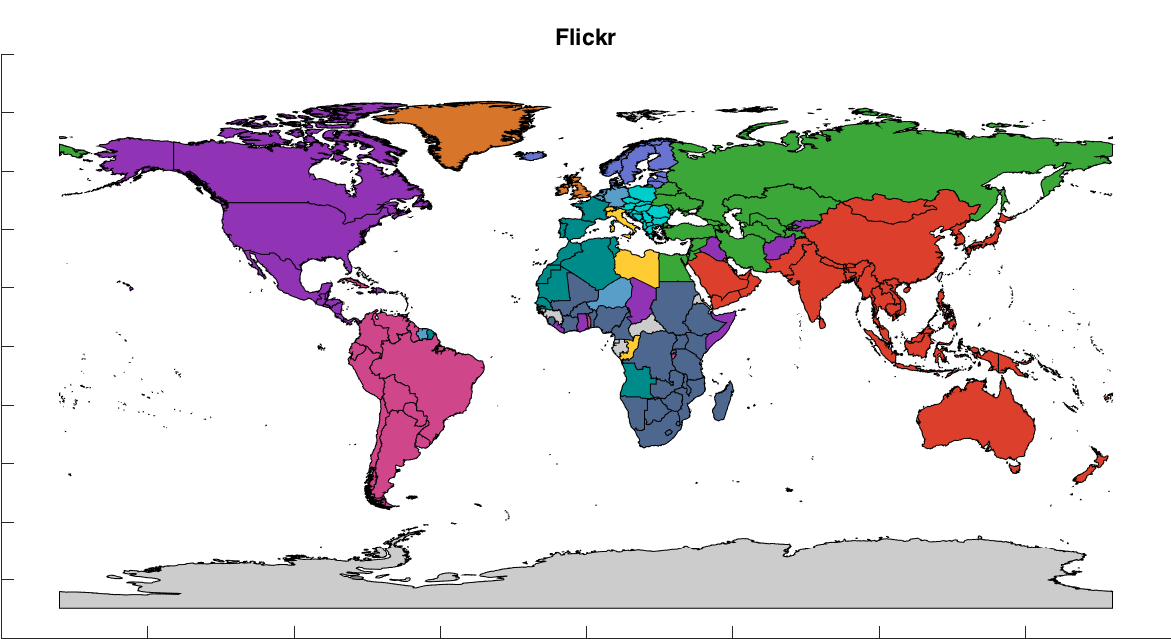}
\includegraphics[width=.49\textwidth]{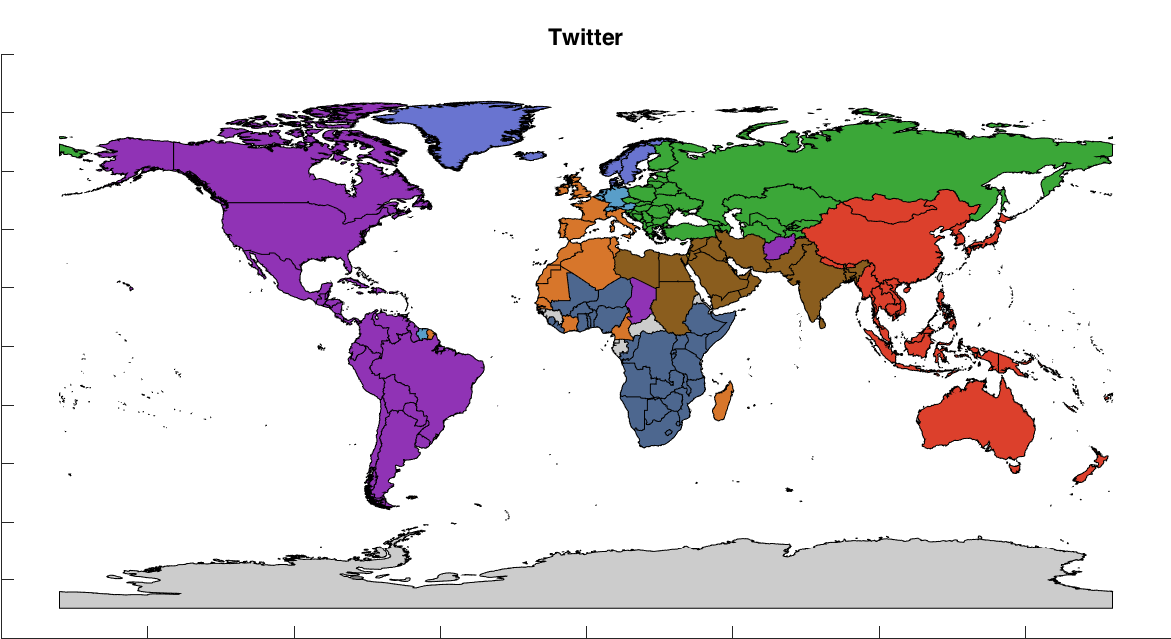}
\includegraphics[width=.49\textwidth]{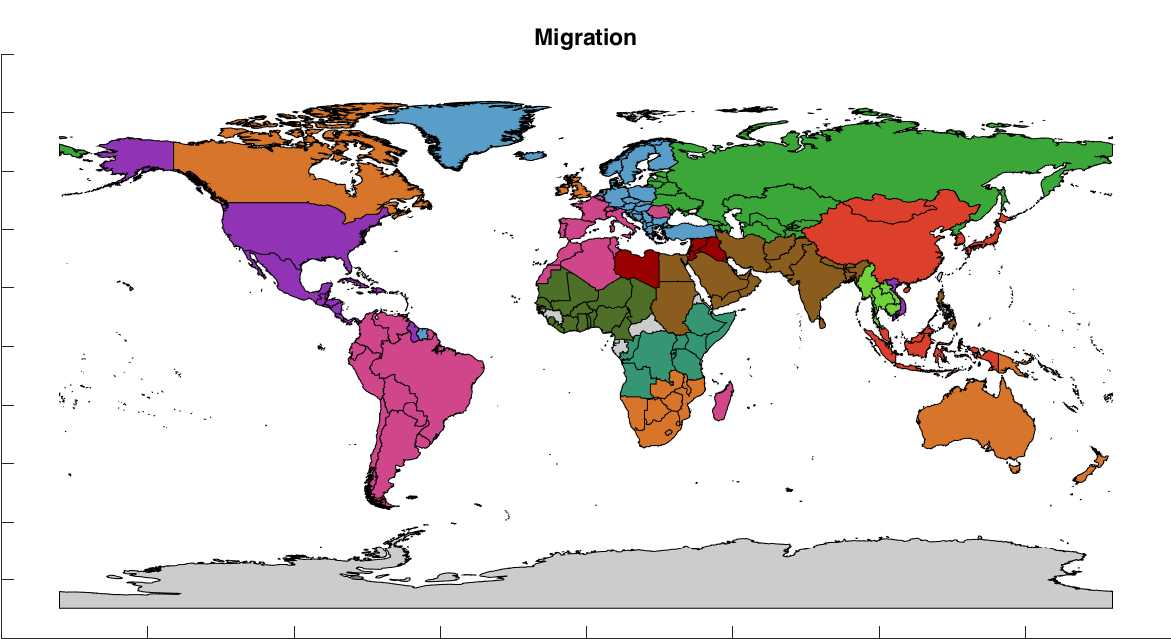}
\includegraphics[width=.49\textwidth]{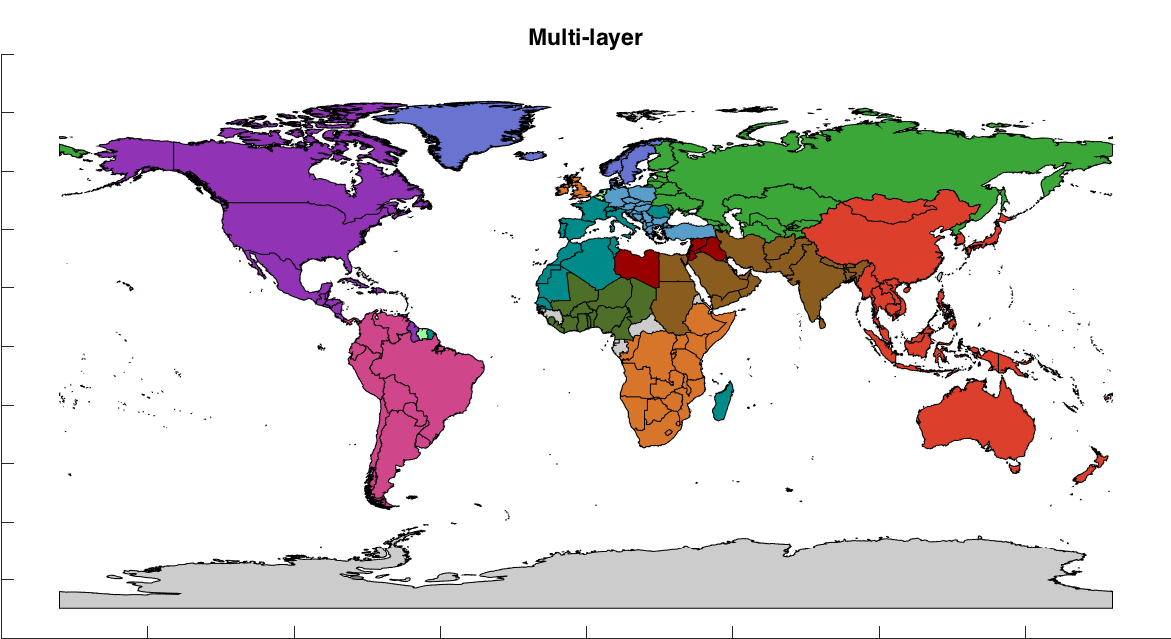}
\caption{\label{fig::partitioning15}Communities for resolution parameter value equal~to~$1.5$.}
\end{figure}

\begin{figure}[h!]
\centering
\includegraphics[width=.49\textwidth]{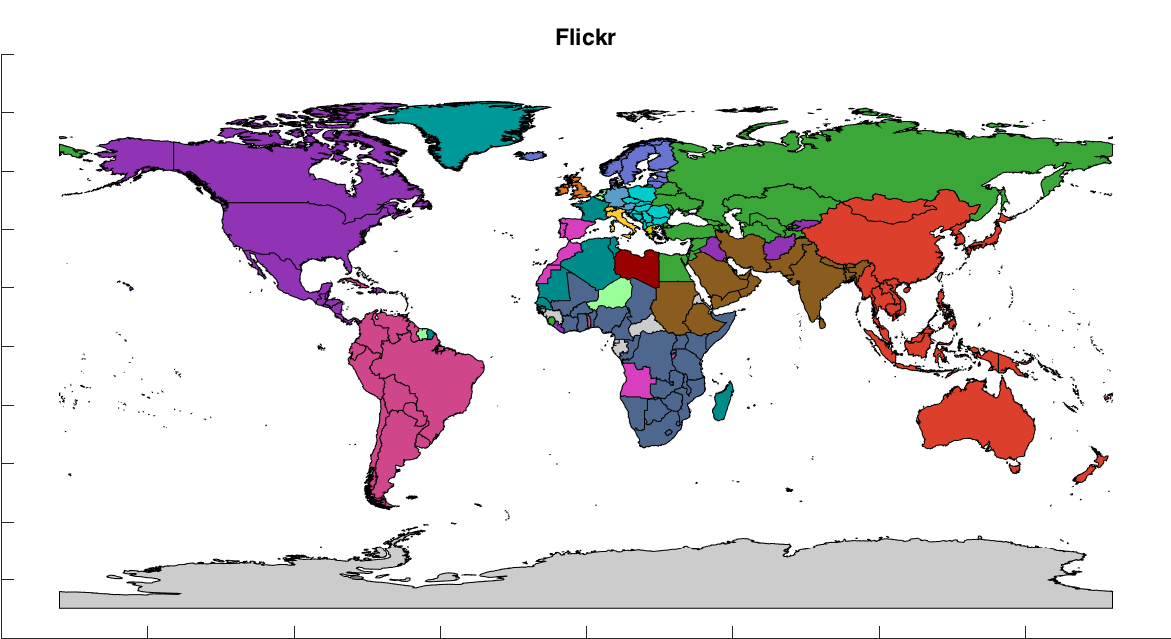}
\includegraphics[width=.49\textwidth]{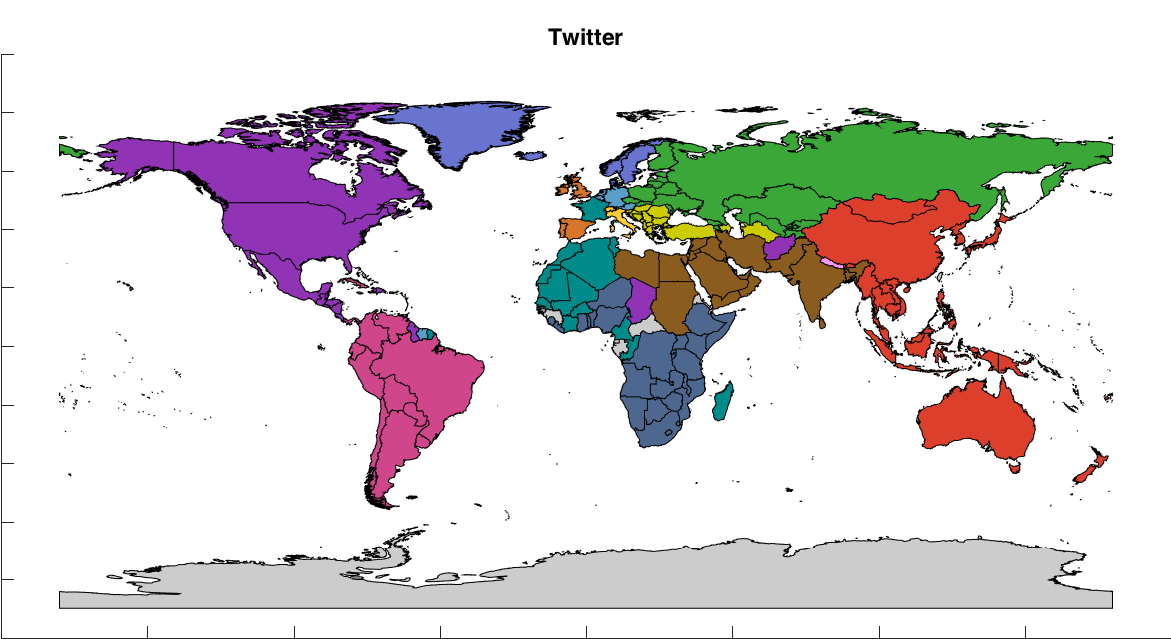}
\includegraphics[width=.49\textwidth]{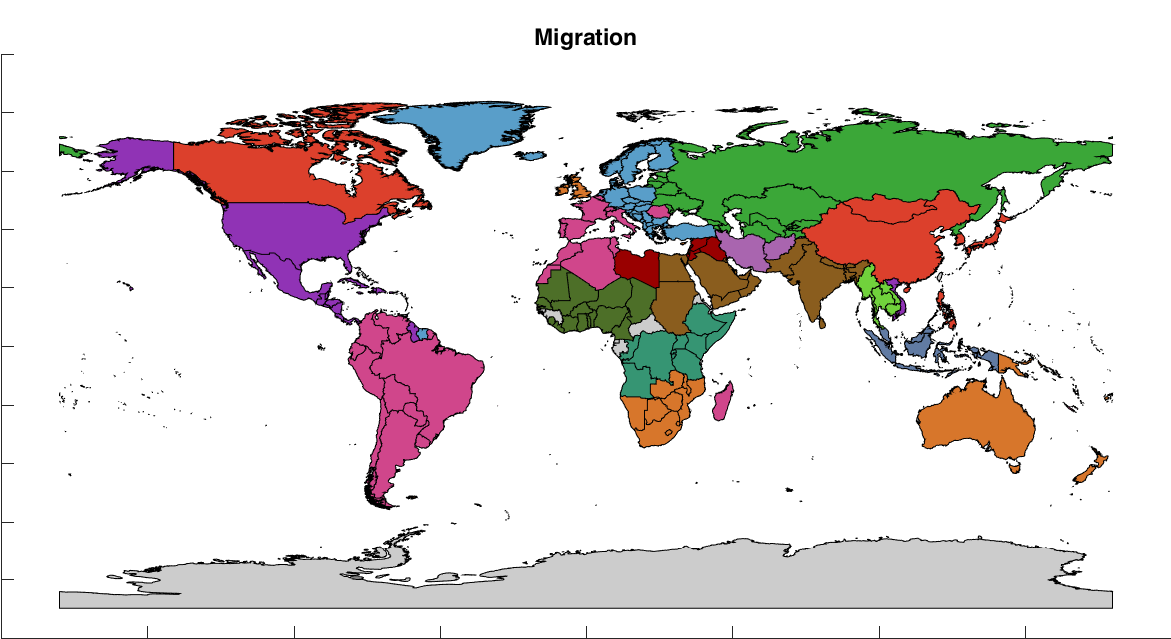}
\includegraphics[width=.49\textwidth]{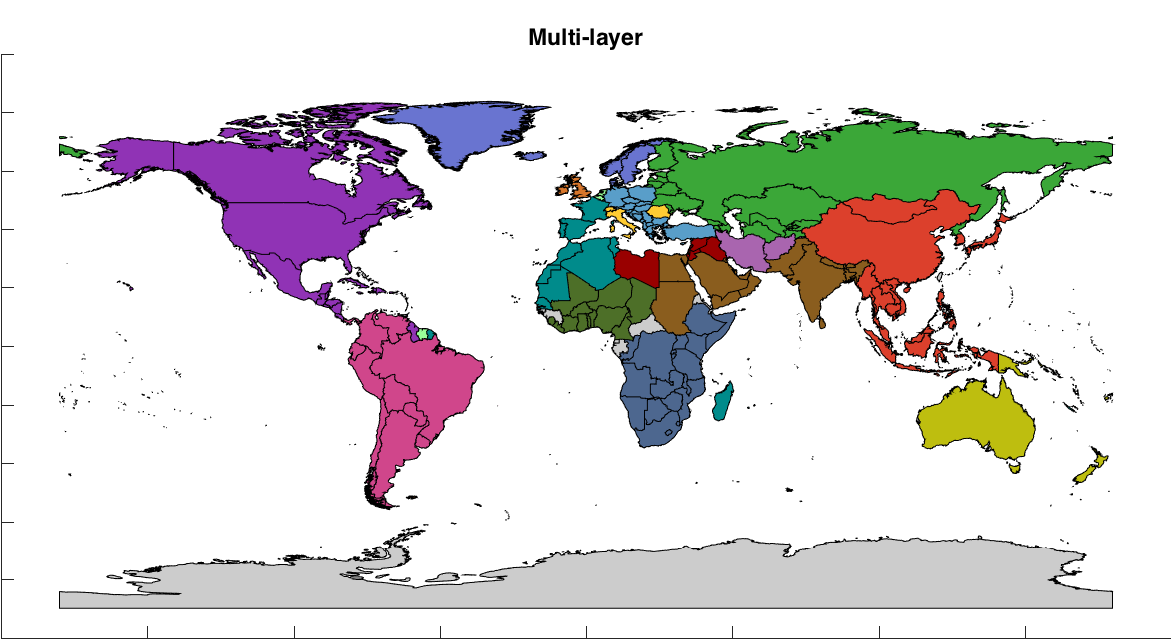}
\caption{\label{fig::partitioning20}Communities for resolution parameter value equal~to~$2.0$.}
\end{figure}

For the sake of noise reduction, we excluded nodes for which incoming or outgoing strength was less than~$10$ at~least in one layer, that left us with a network of $201$ countries. We consider partitioning for different values of resolution parameter for each of the three network layers separately and then for the entire three-layer mobility network. Figure\,\ref{fig::comm_number} shows dependence of the resulting number of communities (provided by the algorithm looking for the optimal partitioning in terms of the adjusted modularity for any number of communities) on the value of a resolution parameter.
When applying modularity maximization with the default resolution parameter of $1.0$ to each layer separately, it leaves us with only \textit{four} and \textit{five} communities for Twitter and Flickr, respectively, while for resolution parameter equal to $2.0$, number of detected communities goes up to \textit{seventeen} making it already harder to visually recognize and analyse different communities on a map. That is why we considered only partitions for parameter taking values between $1.0$ and $2.0$ presenting the results in Figures \ref{fig::partitioning10}, \ref{fig::partitioning15} and \ref{fig::partitioning20} for resolution parameter values $1.0, 1.5$ and $2.0$, respectively. From all figures we can see that partitions of multi-layer network always have less anomalies and are easier to explain than partition of any layer alone.
For example China is often united with North America or even with Canada alone in Flickr and migration layers, which still might be explained, but it feels much more natural to see China as a part of east Asia. In partition of multi-layer network one can always clearly see communities of both Americas (for $a=1$ they united into one community), communities of former USSR countries and Arabic countries. In case of $a=1$ entire Europe is nicely united into one community. Partitions start to be more complicated for higher $a$, but this is because more local patters are discovered. Nevertheless, for multi-layer network Australia is never united with United Kingdom, south-east Asia is always united with the rest of east Asia, south and central Africa form their own communities, i.e., communities are much more geographically cohesive.
Another interesting property of the mobility networks consistent with the previous findings of Ratti~et~al.~\cite{ratti2010redrawing} and Sobolevsky~et~al.~\cite{sobolevsky2013delineating} is that their partitions are spatially connected even though we never imply that. Reasons behind that some countries are grouped into one community can include close geographical distance, strong economical ties and cultural aspects. And this property is particularly clearly observed for the multi-layered network's partitioning.

The final step of providing the evidence for our hypothesis is to evaluate the obtained partitions. For that purpose we compare them with the partitions got from other available types of international connections. As it was previously described in Section~\ref{sec:networks} we chose language, colonial and trade networks for our analysis. We quantify the similarity between partitions using the normalized mutual information (NMI)\,\cite{Danon2005}, which came from information theory and now is also widely used in community detection for partition comparison. After expansion of all expressions for entropy in its definition, NMI~of~two~partitions $A$ and $B$ can be calculated as:
\[
	NMI = 
	\frac{
		-2\sum_{i=1}^{C_A}\sum_{j=1}^{C_B} N_{ij} \log\left( \frac{N_{ij} N}{N^{A}_i N^{B}_j} \right)
	}{
		\sum_{i=1}^{C_A} N^{A}_i \log\left( \frac{N^{A}_i}{N} \right) + \sum_{j=1}^{C_B} N^{B}_j \log\left( \frac{N^{B}_j}{N} \right)
	},
\]
where $C_A$ and $C_B$ are the numbers of communities in each partition, $N^{A}_i$ and $N^{B}_j$ are the cardinalities of each community, $N_{ij}$ are the numbers of nodes classified to community~$i$ in partition~$A$ and to community~$j$ in partition~$B$, $N$ denotes the total number of nodes. NMI~takes values from~$0$ to~$1$ and the higher its value is, the more similar the partitions are, meaning that for identical partitions NMI equals to~$1$.

\begin{figure}[t!]
\centering
\includegraphics[width=.65\textwidth]{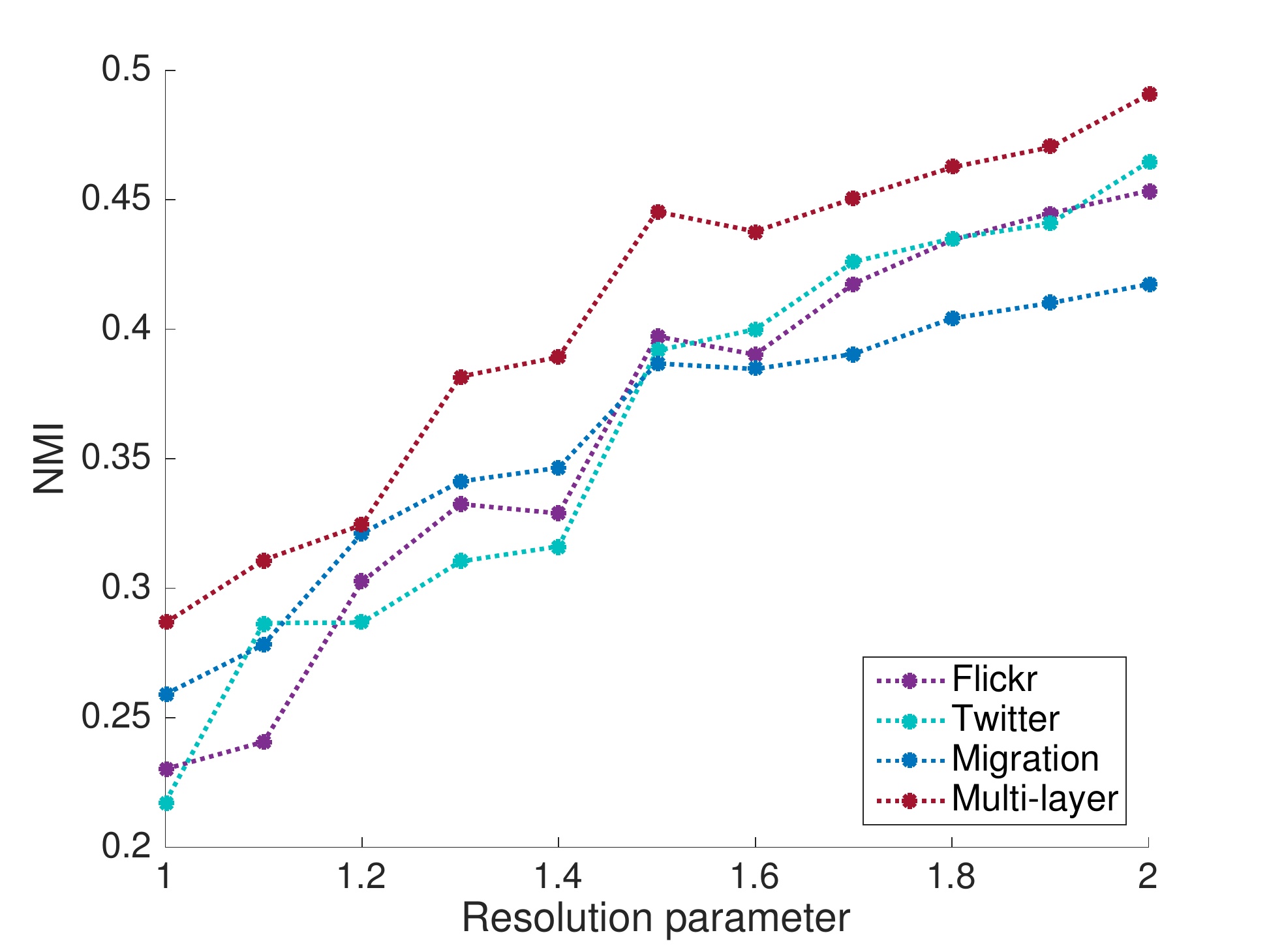}
\caption{\label{fig::NMI}Similarity of community structure between networks of human mobility and other existing international connections.}
\end{figure}

Finally, we compared partitions for each layer and the entire multi-layer network with three partitions of the other networks, quantified similarity of partitions and took average NMI to evaluate how consistent the partitions of each layer and the entire multi-layer network are with the patterns from cultural, historic and economic networks. Results of comparison presented in Figure\,\ref{fig::NMI} show that the community structure of the three-layer mobility network is consistently more similar to the community structure of cultural, historic and economic networks, than community structure of each layer considered separately. This can serve as a good quantitative initial validation of our hypothesis: when considering different aspects of mobility together in a form of the the three-layer mobility network, one can indeed reveal some patterns consistent with other observations better than it could be done by considering any of mobility layers alone.

\section*{Conclusions}
\label{sec:conclusion}

In this study we explored records of different types of human mobility: long-term and short-term. We analysed their similarities and differences and showed that country attractiveness, which is represented with the number of foreign people visiting it, in all three networks follows a log-normal distribution. Moreover, our results showed that normalized weights of links also follow the same distribution with almost the same scaling parameters for Flickr and Twitter networks and much more diverse for the migration one. The existence of a fewer number of stronger migration flows covering majority of the entire migration activity denotes that people tend to move to rather few major countries of interest, while from touristic or other short-term perspective the destinations are more diverse.

We ranked all countries according to their attractiveness for long-term and short-term visits and specifically investigated outlier countries that are highly ranked from one perspective and not from the other. These countries fall into two easily distinguished groups: the first one mostly consist of Arabian countries that could be seen as very attractive for immigrants, but not that much for tourists; and the second one composed of developing countries with beautiful nature and quite high population density which attract a lot of tourists but a fewer amount of migrants. Nevertheless, there is also a group of highly developed countries that attract both tourists and immigrants.

We also looked at how different types of mobility could be predicted by existing models. After fitting the gravity model to all layers of our multi-layer mobility network, we showed that long-term mobility is much more local than short-term one meaning that when people choose where to live they take into an account distance much more often than when choosing where to go for a trip.

Finally, as we found that different datasets provided different perspectives on human mobility, we combined them into one three-layer mobility network. We showed that considering all three mobility networks together as one single multi-layer network, helps us to better describe the structure of the global human society in a way which is more consistent with other types of known links between the countries. Namely, we applied a community detection method to three networks separately together with the multi-layer one, and compared all four resulting partitions with those obtained from networks of other existing international connections (i.e., language similarity, colonial relations and international trade). The results showed that the three-layer mobility network provides an underlying structure consistently more similar to the structures behind those international connection networks, compared to each of the three mobility network layers taken separately. We also discuss the specific spatial patterns revealed. Therefore, our general conclusion is that considering human mobility network from a multi-layer perspective is crucial as multi-layer mobility networks can reveal some important patterns which single networks cannot. 

\section*{Acknowledgments}
The authors thank Accenture China, American Air Liquide, BBVA, Emirates Integrated Telecommunications Company (du), ENEL Foundation, Ericsson, Kuwait--MIT Center for Natural Resources and the Environment, Liberty Mutual Institute, Singapore--MIT Alliance for Research and Technology (SMART), Regional Municipality of Wood Buffalo, Volkswagen Electronics Research Lab, and all the members of the MIT Senseable City Lab Consortium for supporting this research. Moreover, the authors would like to acknowledge the Austrian Science Fund (FWF) through the Doctoral College GIScience (DK W 1237--N23), Department of Geoinformatics~-- Z\_GIS, University of Salzburg, Austria. Finally, a part of this research was also supported by the research project `Managing Trust and Coordinating Interactions in Smart Networks of People, Machines and Organizations' which is funded by the Croatian Science Foundation.

\bibliographystyle{apsrev}
\bibliography{literature}

\begin{thebibliography}{57}
\expandafter\ifx\csname natexlab\endcsname\relax\def\natexlab#1{#1}\fi
\expandafter\ifx\csname bibnamefont\endcsname\relax
  \def\bibnamefont#1{#1}\fi
\expandafter\ifx\csname bibfnamefont\endcsname\relax
  \def\bibfnamefont#1{#1}\fi
\expandafter\ifx\csname citenamefont\endcsname\relax
  \def\citenamefont#1{#1}\fi
\expandafter\ifx\csname url\endcsname\relax
  \def\url#1{\texttt{#1}}\fi
\expandafter\ifx\csname urlprefix\endcsname\relax\def\urlprefix{URL }\fi
\providecommand{\bibinfo}[2]{#2}
\providecommand{\eprint}[2][]{\url{#2}}

\bibitem[{\citenamefont{Ratti et~al.}(2006)\citenamefont{Ratti, Williams,
  Frenchman, and Pulselli}}]{ratti2006mlu}
\bibinfo{author}{\bibfnamefont{C.}~\bibnamefont{Ratti}},
  \bibinfo{author}{\bibfnamefont{S.}~\bibnamefont{Williams}},
  \bibinfo{author}{\bibfnamefont{D.}~\bibnamefont{Frenchman}},
  \bibnamefont{and} \bibinfo{author}{\bibfnamefont{R.}~\bibnamefont{Pulselli}},
  \bibinfo{journal}{Environment and Planning B} \textbf{\bibinfo{volume}{33}},
  \bibinfo{pages}{727} (\bibinfo{year}{2006}).

\bibitem[{\citenamefont{Calabrese and Ratti}(2006)}]{calabrese2006real}
\bibinfo{author}{\bibfnamefont{F.}~\bibnamefont{Calabrese}} \bibnamefont{and}
  \bibinfo{author}{\bibfnamefont{C.}~\bibnamefont{Ratti}},
  \bibinfo{journal}{Networks and Communication Studies}
  \textbf{\bibinfo{volume}{20}}, \bibinfo{pages}{247} (\bibinfo{year}{2006}).

\bibitem[{\citenamefont{Girardin et~al.}(2008)\citenamefont{Girardin,
  Calabrese, Fiore, Ratti, and Blat}}]{girardin2008digital}
\bibinfo{author}{\bibfnamefont{F.}~\bibnamefont{Girardin}},
  \bibinfo{author}{\bibfnamefont{F.}~\bibnamefont{Calabrese}},
  \bibinfo{author}{\bibfnamefont{F.~D.} \bibnamefont{Fiore}},
  \bibinfo{author}{\bibfnamefont{C.}~\bibnamefont{Ratti}}, \bibnamefont{and}
  \bibinfo{author}{\bibfnamefont{J.}~\bibnamefont{Blat}},
  \bibinfo{journal}{IEEE Pervasive Computing} \textbf{\bibinfo{volume}{7}},
  \bibinfo{pages}{36} (\bibinfo{year}{2008}).

\bibitem[{\citenamefont{Quercia et~al.}(2010)\citenamefont{Quercia, Lathia,
  Calabrese, Di~Lorenzo, and Crowcroft}}]{quercia2010rse}
\bibinfo{author}{\bibfnamefont{D.}~\bibnamefont{Quercia}},
  \bibinfo{author}{\bibfnamefont{N.}~\bibnamefont{Lathia}},
  \bibinfo{author}{\bibfnamefont{F.}~\bibnamefont{Calabrese}},
  \bibinfo{author}{\bibfnamefont{G.}~\bibnamefont{Di~Lorenzo}},
  \bibnamefont{and}
  \bibinfo{author}{\bibfnamefont{J.}~\bibnamefont{Crowcroft}}, in
  \emph{\bibinfo{booktitle}{Proceedings of the 10th IEEE International
  Conference on Data Mining}} (\bibinfo{publisher}{IEEE},
  \bibinfo{year}{2010}), pp. \bibinfo{pages}{971--976}.

\bibitem[{\citenamefont{Santi et~al.}(2014)\citenamefont{Santi, Resta, Szell,
  Sobolevsky, Strogatz, and Ratti}}]{santi2013taxi}
\bibinfo{author}{\bibfnamefont{P.}~\bibnamefont{Santi}},
  \bibinfo{author}{\bibfnamefont{G.}~\bibnamefont{Resta}},
  \bibinfo{author}{\bibfnamefont{M.}~\bibnamefont{Szell}},
  \bibinfo{author}{\bibfnamefont{S.}~\bibnamefont{Sobolevsky}},
  \bibinfo{author}{\bibfnamefont{S.}~\bibnamefont{Strogatz}}, \bibnamefont{and}
  \bibinfo{author}{\bibfnamefont{C.}~\bibnamefont{Ratti}},
  \bibinfo{journal}{Proceedings of the National Academy of Sciences}
  \textbf{\bibinfo{volume}{111}}, \bibinfo{pages}{13290}
  (\bibinfo{year}{2014}).

\bibitem[{\citenamefont{Kang et~al.}(2013)\citenamefont{Kang, Sobolevsky, Liu,
  and Ratti}}]{kang2013exploring}
\bibinfo{author}{\bibfnamefont{C.}~\bibnamefont{Kang}},
  \bibinfo{author}{\bibfnamefont{S.}~\bibnamefont{Sobolevsky}},
  \bibinfo{author}{\bibfnamefont{Y.}~\bibnamefont{Liu}}, \bibnamefont{and}
  \bibinfo{author}{\bibfnamefont{C.}~\bibnamefont{Ratti}}, in
  \emph{\bibinfo{booktitle}{Proceedings of the 2nd ACM SIGKDD International
  Workshop on Urban Computing}} (\bibinfo{publisher}{ACM New York, NY, USA},
  \bibinfo{year}{2013}), pp. \bibinfo{pages}{1--8}.

\bibitem[{\citenamefont{Bagchi and White}(2005)}]{bagchi2005}
\bibinfo{author}{\bibfnamefont{M.}~\bibnamefont{Bagchi}} \bibnamefont{and}
  \bibinfo{author}{\bibfnamefont{P.}~\bibnamefont{White}},
  \bibinfo{journal}{Transport Policy} \textbf{\bibinfo{volume}{12}},
  \bibinfo{pages}{464} (\bibinfo{year}{2005}).

\bibitem[{\citenamefont{Lathia et~al.}(2012)\citenamefont{Lathia, Quercia, and
  Crowcroft}}]{lathia2012}
\bibinfo{author}{\bibfnamefont{N.}~\bibnamefont{Lathia}},
  \bibinfo{author}{\bibfnamefont{D.}~\bibnamefont{Quercia}}, \bibnamefont{and}
  \bibinfo{author}{\bibfnamefont{J.}~\bibnamefont{Crowcroft}}, in
  \emph{\bibinfo{booktitle}{Pervasive Computing}}, edited by
  \bibinfo{editor}{\bibfnamefont{J.}~\bibnamefont{Kay}},
  \bibinfo{editor}{\bibfnamefont{P.}~\bibnamefont{Lukowicz}},
  \bibinfo{editor}{\bibfnamefont{H.}~\bibnamefont{Tokuda}},
  \bibinfo{editor}{\bibfnamefont{P.}~\bibnamefont{Olivier}}, \bibnamefont{and}
  \bibinfo{editor}{\bibfnamefont{A.}~\bibnamefont{Krüger}}
  (\bibinfo{publisher}{Springer}, \bibinfo{year}{2012}), vol.
  \bibinfo{volume}{7319} of \emph{\bibinfo{series}{Lecture Notes in Computer
  Science}}, pp. \bibinfo{pages}{91--98}.

\bibitem[{\citenamefont{Java et~al.}(2007)\citenamefont{Java, Song, Finin, and
  Tseng}}]{java2007we}
\bibinfo{author}{\bibfnamefont{A.}~\bibnamefont{Java}},
  \bibinfo{author}{\bibfnamefont{X.}~\bibnamefont{Song}},
  \bibinfo{author}{\bibfnamefont{T.}~\bibnamefont{Finin}}, \bibnamefont{and}
  \bibinfo{author}{\bibfnamefont{B.}~\bibnamefont{Tseng}}, in
  \emph{\bibinfo{booktitle}{Proceedings of the 9th WebKDD and 1st SNA-KDD 2007
  workshop on Web mining and social network analysis}} (\bibinfo{publisher}{ACM
  New York, NY, USA}, \bibinfo{year}{2007}), pp. \bibinfo{pages}{56--65}.

\bibitem[{\citenamefont{Szell et~al.}(2014)\citenamefont{Szell, Grauwin, and
  Ratti}}]{szell2013}
\bibinfo{author}{\bibfnamefont{M.}~\bibnamefont{Szell}},
  \bibinfo{author}{\bibfnamefont{S.}~\bibnamefont{Grauwin}}, \bibnamefont{and}
  \bibinfo{author}{\bibfnamefont{C.}~\bibnamefont{Ratti}},
  \bibinfo{journal}{PloS One} \textbf{\bibinfo{volume}{9}}, \bibinfo{pages}{1}
  (\bibinfo{year}{2014}).

\bibitem[{\citenamefont{Frank et~al.}(2013)\citenamefont{Frank, Mitchell,
  Dodds, and Danforth}}]{frank2013happiness}
\bibinfo{author}{\bibfnamefont{M.~R.} \bibnamefont{Frank}},
  \bibinfo{author}{\bibfnamefont{L.}~\bibnamefont{Mitchell}},
  \bibinfo{author}{\bibfnamefont{P.~S.} \bibnamefont{Dodds}}, \bibnamefont{and}
  \bibinfo{author}{\bibfnamefont{C.~M.} \bibnamefont{Danforth}},
  \bibinfo{journal}{Scientific Reports} pp. \bibinfo{pages}{1--9}
  (\bibinfo{year}{2013}).

\bibitem[{\citenamefont{Sobolevsky
  et~al.}(2014{\natexlab{a}})\citenamefont{Sobolevsky, Sitko, Grauwin,
  Tachet~des Combes, Hawelka, Arias, and Ratti}}]{sobolevsky2014mining}
\bibinfo{author}{\bibfnamefont{S.}~\bibnamefont{Sobolevsky}},
  \bibinfo{author}{\bibfnamefont{I.}~\bibnamefont{Sitko}},
  \bibinfo{author}{\bibfnamefont{S.}~\bibnamefont{Grauwin}},
  \bibinfo{author}{\bibfnamefont{R.}~\bibnamefont{Tachet~des Combes}},
  \bibinfo{author}{\bibfnamefont{B.}~\bibnamefont{Hawelka}},
  \bibinfo{author}{\bibfnamefont{J.~M.} \bibnamefont{Arias}}, \bibnamefont{and}
  \bibinfo{author}{\bibfnamefont{C.}~\bibnamefont{Ratti}}, in
  \emph{\bibinfo{booktitle}{Proceedings of the 2nd ASE International Conference
  on Big data Science and Computing}} (\bibinfo{publisher}{ASE, Stanford},
  \bibinfo{year}{2014}{\natexlab{a}}), pp. \bibinfo{pages}{1--10}.

\bibitem[{\citenamefont{Sobolevsky
  et~al.}(2014{\natexlab{b}})\citenamefont{Sobolevsky, Sitko, Tachet~des
  Combes, Hawelka, Arias, and Ratti}}]{sobolevsky2014money}
\bibinfo{author}{\bibfnamefont{S.}~\bibnamefont{Sobolevsky}},
  \bibinfo{author}{\bibfnamefont{I.}~\bibnamefont{Sitko}},
  \bibinfo{author}{\bibfnamefont{R.}~\bibnamefont{Tachet~des Combes}},
  \bibinfo{author}{\bibfnamefont{B.}~\bibnamefont{Hawelka}},
  \bibinfo{author}{\bibfnamefont{J.~M.} \bibnamefont{Arias}}, \bibnamefont{and}
  \bibinfo{author}{\bibfnamefont{C.}~\bibnamefont{Ratti}}, in
  \emph{\bibinfo{booktitle}{Proceedings of the IEEE International Congress on
  Big Data}} (\bibinfo{publisher}{IEEE}, \bibinfo{year}{2014}{\natexlab{b}}),
  pp. \bibinfo{pages}{136--143}.

\bibitem[{\citenamefont{Sobolevsky
  et~al.}(2015{\natexlab{a}})\citenamefont{Sobolevsky, Sitko, Combes, Hawelka,
  Arias, and Ratti}}]{sobolevsky2015cities}
\bibinfo{author}{\bibfnamefont{S.}~\bibnamefont{Sobolevsky}},
  \bibinfo{author}{\bibfnamefont{I.}~\bibnamefont{Sitko}},
  \bibinfo{author}{\bibfnamefont{R.~T.~d.} \bibnamefont{Combes}},
  \bibinfo{author}{\bibfnamefont{B.}~\bibnamefont{Hawelka}},
  \bibinfo{author}{\bibfnamefont{J.~M.} \bibnamefont{Arias}}, \bibnamefont{and}
  \bibinfo{author}{\bibfnamefont{C.}~\bibnamefont{Ratti}},
  \bibinfo{journal}{arXiv preprint arXiv:1505.03854} pp. \bibinfo{pages}{1--21}
  (\bibinfo{year}{2015}{\natexlab{a}}).

\bibitem[{\citenamefont{Sobolevsky
  et~al.}(2015{\natexlab{b}})\citenamefont{Sobolevsky, Massaro, Bojic, Arias,
  and Ratti}}]{sobolevsky2015predicting}
\bibinfo{author}{\bibfnamefont{S.}~\bibnamefont{Sobolevsky}},
  \bibinfo{author}{\bibfnamefont{E.}~\bibnamefont{Massaro}},
  \bibinfo{author}{\bibfnamefont{I.}~\bibnamefont{Bojic}},
  \bibinfo{author}{\bibfnamefont{J.~M.} \bibnamefont{Arias}}, \bibnamefont{and}
  \bibinfo{author}{\bibfnamefont{C.}~\bibnamefont{Ratti}}, in
  \emph{\bibinfo{booktitle}{Proceedings of the 6th ASE International Conference
  on Data Science}} (\bibinfo{publisher}{ASE, Stanford},
  \bibinfo{year}{2015}{\natexlab{b}}), pp. \bibinfo{pages}{1--12}.

\bibitem[{\citenamefont{Ratti et~al.}(2010)\citenamefont{Ratti, Sobolevsky,
  Calabrese, Andris, Reades, Martino, Claxton, and
  Strogatz}}]{ratti2010redrawing}
\bibinfo{author}{\bibfnamefont{C.}~\bibnamefont{Ratti}},
  \bibinfo{author}{\bibfnamefont{S.}~\bibnamefont{Sobolevsky}},
  \bibinfo{author}{\bibfnamefont{F.}~\bibnamefont{Calabrese}},
  \bibinfo{author}{\bibfnamefont{C.}~\bibnamefont{Andris}},
  \bibinfo{author}{\bibfnamefont{J.}~\bibnamefont{Reades}},
  \bibinfo{author}{\bibfnamefont{M.}~\bibnamefont{Martino}},
  \bibinfo{author}{\bibfnamefont{R.}~\bibnamefont{Claxton}}, \bibnamefont{and}
  \bibinfo{author}{\bibfnamefont{S.~H.} \bibnamefont{Strogatz}},
  \bibinfo{journal}{PLoS One} \textbf{\bibinfo{volume}{5}}, \bibinfo{pages}{1}
  (\bibinfo{year}{2010}).

\bibitem[{\citenamefont{Sobolevsky et~al.}(2013)\citenamefont{Sobolevsky,
  Szell, Campari, Couronn{\'e}, Smoreda, and
  Ratti}}]{sobolevsky2013delineating}
\bibinfo{author}{\bibfnamefont{S.}~\bibnamefont{Sobolevsky}},
  \bibinfo{author}{\bibfnamefont{M.}~\bibnamefont{Szell}},
  \bibinfo{author}{\bibfnamefont{R.}~\bibnamefont{Campari}},
  \bibinfo{author}{\bibfnamefont{T.}~\bibnamefont{Couronn{\'e}}},
  \bibinfo{author}{\bibfnamefont{Z.}~\bibnamefont{Smoreda}}, \bibnamefont{and}
  \bibinfo{author}{\bibfnamefont{C.}~\bibnamefont{Ratti}},
  \bibinfo{journal}{PloS One} \textbf{\bibinfo{volume}{8}}, \bibinfo{pages}{1}
  (\bibinfo{year}{2013}).

\bibitem[{\citenamefont{Pei et~al.}(2014)\citenamefont{Pei, Sobolevsky, Ratti,
  Shaw, Li, and Zhou}}]{pei2014new}
\bibinfo{author}{\bibfnamefont{T.}~\bibnamefont{Pei}},
  \bibinfo{author}{\bibfnamefont{S.}~\bibnamefont{Sobolevsky}},
  \bibinfo{author}{\bibfnamefont{C.}~\bibnamefont{Ratti}},
  \bibinfo{author}{\bibfnamefont{S.-L.} \bibnamefont{Shaw}},
  \bibinfo{author}{\bibfnamefont{T.}~\bibnamefont{Li}}, \bibnamefont{and}
  \bibinfo{author}{\bibfnamefont{C.}~\bibnamefont{Zhou}},
  \bibinfo{journal}{International Journal of Geographical Information Science}
  \textbf{\bibinfo{volume}{28}}, \bibinfo{pages}{1988} (\bibinfo{year}{2014}).

\bibitem[{\citenamefont{Grauwin
  et~al.}(2015{\natexlab{a}})\citenamefont{Grauwin, Sobolevsky, Moritz,
  G{\'o}dor, and Ratti}}]{grauwin2014towards}
\bibinfo{author}{\bibfnamefont{S.}~\bibnamefont{Grauwin}},
  \bibinfo{author}{\bibfnamefont{S.}~\bibnamefont{Sobolevsky}},
  \bibinfo{author}{\bibfnamefont{S.}~\bibnamefont{Moritz}},
  \bibinfo{author}{\bibfnamefont{I.}~\bibnamefont{G{\'o}dor}},
  \bibnamefont{and} \bibinfo{author}{\bibfnamefont{C.}~\bibnamefont{Ratti}}, in
  \emph{\bibinfo{booktitle}{Computational Approaches for Urban Environments}},
  edited by \bibinfo{editor}{\bibfnamefont{M.}~\bibnamefont{Helbich}},
  \bibinfo{editor}{\bibfnamefont{J.~J.} \bibnamefont{Arsanjani}},
  \bibnamefont{and} \bibinfo{editor}{\bibfnamefont{M.}~\bibnamefont{Leitner}}
  (\bibinfo{publisher}{Springer}, \bibinfo{year}{2015}{\natexlab{a}}),
  vol.~\bibinfo{volume}{13} of \emph{\bibinfo{series}{Geotechnologies and the
  Environment}}, pp. \bibinfo{pages}{363--387}.

\bibitem[{\citenamefont{Gonz{\'a}lez et~al.}(2008)\citenamefont{Gonz{\'a}lez,
  Hidalgo, and Barab{\'a}si}}]{gonzalez2008uih}
\bibinfo{author}{\bibfnamefont{M.}~\bibnamefont{Gonz{\'a}lez}},
  \bibinfo{author}{\bibfnamefont{C.}~\bibnamefont{Hidalgo}}, \bibnamefont{and}
  \bibinfo{author}{\bibfnamefont{A.-L.} \bibnamefont{Barab{\'a}si}},
  \bibinfo{journal}{Nature} \textbf{\bibinfo{volume}{453}},
  \bibinfo{pages}{779} (\bibinfo{year}{2008}).

\bibitem[{\citenamefont{Kung et~al.}(2014)\citenamefont{Kung, Greco,
  Sobolevsky, and Ratti}}]{kung2014exploring}
\bibinfo{author}{\bibfnamefont{K.}~\bibnamefont{Kung}},
  \bibinfo{author}{\bibfnamefont{K.}~\bibnamefont{Greco}},
  \bibinfo{author}{\bibfnamefont{S.}~\bibnamefont{Sobolevsky}},
  \bibnamefont{and} \bibinfo{author}{\bibfnamefont{C.}~\bibnamefont{Ratti}},
  \bibinfo{journal}{PLoS One} \textbf{\bibinfo{volume}{9}}, \bibinfo{pages}{1}
  (\bibinfo{year}{2014}).

\bibitem[{\citenamefont{Hoteit et~al.}(2014)\citenamefont{Hoteit, Secci,
  Sobolevsky, Ratti, and Pujolle}}]{hoteit2014estimating}
\bibinfo{author}{\bibfnamefont{S.}~\bibnamefont{Hoteit}},
  \bibinfo{author}{\bibfnamefont{S.}~\bibnamefont{Secci}},
  \bibinfo{author}{\bibfnamefont{S.}~\bibnamefont{Sobolevsky}},
  \bibinfo{author}{\bibfnamefont{C.}~\bibnamefont{Ratti}}, \bibnamefont{and}
  \bibinfo{author}{\bibfnamefont{G.}~\bibnamefont{Pujolle}},
  \bibinfo{journal}{Computer Networks} \textbf{\bibinfo{volume}{64}},
  \bibinfo{pages}{296} (\bibinfo{year}{2014}).

\bibitem[{\citenamefont{Amini et~al.}(2014)\citenamefont{Amini, Kung, Kang,
  Sobolevsky, and Ratti}}]{amini2014impact}
\bibinfo{author}{\bibfnamefont{A.}~\bibnamefont{Amini}},
  \bibinfo{author}{\bibfnamefont{K.}~\bibnamefont{Kung}},
  \bibinfo{author}{\bibfnamefont{C.}~\bibnamefont{Kang}},
  \bibinfo{author}{\bibfnamefont{S.}~\bibnamefont{Sobolevsky}},
  \bibnamefont{and} \bibinfo{author}{\bibfnamefont{C.}~\bibnamefont{Ratti}},
  \bibinfo{journal}{EPJ Data Science} \textbf{\bibinfo{volume}{3}},
  \bibinfo{pages}{1} (\bibinfo{year}{2014}).

\bibitem[{\citenamefont{Greenwood}(1985)}]{greenwood1985human}
\bibinfo{author}{\bibfnamefont{M.~J.} \bibnamefont{Greenwood}},
  \bibinfo{journal}{Journal of regional Science} \textbf{\bibinfo{volume}{25}},
  \bibinfo{pages}{521} (\bibinfo{year}{1985}).

\bibitem[{\citenamefont{Fagiolo and
  Mastrorillo}(2013)}]{fagiolo2013international}
\bibinfo{author}{\bibfnamefont{G.}~\bibnamefont{Fagiolo}} \bibnamefont{and}
  \bibinfo{author}{\bibfnamefont{M.}~\bibnamefont{Mastrorillo}},
  \bibinfo{journal}{Physical Review E} \textbf{\bibinfo{volume}{88}},
  \bibinfo{pages}{012812} (\bibinfo{year}{2013}).

\bibitem[{\citenamefont{Abel and Sander}(2014)}]{abel2014quantifying}
\bibinfo{author}{\bibfnamefont{G.~J.} \bibnamefont{Abel}} \bibnamefont{and}
  \bibinfo{author}{\bibfnamefont{N.}~\bibnamefont{Sander}},
  \bibinfo{journal}{Science} \textbf{\bibinfo{volume}{343}},
  \bibinfo{pages}{1520} (\bibinfo{year}{2014}).

\bibitem[{\citenamefont{Tranos et~al.}(2015)\citenamefont{Tranos, Gheasi, and
  Nijkamp}}]{tranos2012international}
\bibinfo{author}{\bibfnamefont{E.}~\bibnamefont{Tranos}},
  \bibinfo{author}{\bibfnamefont{M.}~\bibnamefont{Gheasi}}, \bibnamefont{and}
  \bibinfo{author}{\bibfnamefont{P.}~\bibnamefont{Nijkamp}},
  \bibinfo{journal}{Environment and Planning B: Planning and Design}
  \textbf{\bibinfo{volume}{42}}, \bibinfo{pages}{4} (\bibinfo{year}{2015}).

\bibitem[{\citenamefont{Hawelka et~al.}(2014)\citenamefont{Hawelka, Sitko,
  Beinat, Sobolevsky, Kazakopoulos, and Ratti}}]{hawelka2014}
\bibinfo{author}{\bibfnamefont{B.}~\bibnamefont{Hawelka}},
  \bibinfo{author}{\bibfnamefont{I.}~\bibnamefont{Sitko}},
  \bibinfo{author}{\bibfnamefont{E.}~\bibnamefont{Beinat}},
  \bibinfo{author}{\bibfnamefont{S.}~\bibnamefont{Sobolevsky}},
  \bibinfo{author}{\bibfnamefont{P.}~\bibnamefont{Kazakopoulos}},
  \bibnamefont{and} \bibinfo{author}{\bibfnamefont{C.}~\bibnamefont{Ratti}},
  \bibinfo{journal}{Cartography and Geographic Information Science}
  \textbf{\bibinfo{volume}{41}}, \bibinfo{pages}{260} (\bibinfo{year}{2014}).

\bibitem[{\citenamefont{Paldino et~al.}(2015)\citenamefont{Paldino, Bojic,
  Sobolevsky, Ratti, and Gonz{\'a}lez}}]{paldino2015flickr}
\bibinfo{author}{\bibfnamefont{S.}~\bibnamefont{Paldino}},
  \bibinfo{author}{\bibfnamefont{I.}~\bibnamefont{Bojic}},
  \bibinfo{author}{\bibfnamefont{S.}~\bibnamefont{Sobolevsky}},
  \bibinfo{author}{\bibfnamefont{C.}~\bibnamefont{Ratti}}, \bibnamefont{and}
  \bibinfo{author}{\bibfnamefont{M.~C.} \bibnamefont{Gonz{\'a}lez}},
  \bibinfo{journal}{EPJ Data Science} \textbf{\bibinfo{volume}{4}},
  \bibinfo{pages}{1} (\bibinfo{year}{2015}).

\bibitem[{\citenamefont{Sobolevsky
  et~al.}(2015{\natexlab{c}})\citenamefont{Sobolevsky, Bojic, Belyi, Sitko,
  Hawelka, Murillo~Arias, and Ratti}}]{sobolevsky2015scaling}
\bibinfo{author}{\bibfnamefont{S.}~\bibnamefont{Sobolevsky}},
  \bibinfo{author}{\bibfnamefont{I.}~\bibnamefont{Bojic}},
  \bibinfo{author}{\bibfnamefont{A.}~\bibnamefont{Belyi}},
  \bibinfo{author}{\bibfnamefont{I.}~\bibnamefont{Sitko}},
  \bibinfo{author}{\bibfnamefont{B.}~\bibnamefont{Hawelka}},
  \bibinfo{author}{\bibfnamefont{J.}~\bibnamefont{Murillo~Arias}},
  \bibnamefont{and} \bibinfo{author}{\bibfnamefont{C.}~\bibnamefont{Ratti}}, in
  \emph{\bibinfo{booktitle}{Big Data (BigData Congress), 2015 IEEE
  International Congress on}} (\bibinfo{publisher}{IEEE},
  \bibinfo{year}{2015}{\natexlab{c}}), pp. \bibinfo{pages}{600--607}.

\bibitem[{\citenamefont{Sgrignoli et~al.}(2015)\citenamefont{Sgrignoli,
  Metulini, Schiavo, and Riccaboni}}]{sgrignoli2015relation}
\bibinfo{author}{\bibfnamefont{P.}~\bibnamefont{Sgrignoli}},
  \bibinfo{author}{\bibfnamefont{R.}~\bibnamefont{Metulini}},
  \bibinfo{author}{\bibfnamefont{S.}~\bibnamefont{Schiavo}}, \bibnamefont{and}
  \bibinfo{author}{\bibfnamefont{M.}~\bibnamefont{Riccaboni}},
  \bibinfo{journal}{Physica A: Statistical Mechanics and its Applications}
  \textbf{\bibinfo{volume}{417}}, \bibinfo{pages}{245} (\bibinfo{year}{2015}).

\bibitem[{\citenamefont{Fagiolo and Mastrorillo}(2014)}]{fagiolo2014does}
\bibinfo{author}{\bibfnamefont{G.}~\bibnamefont{Fagiolo}} \bibnamefont{and}
  \bibinfo{author}{\bibfnamefont{M.}~\bibnamefont{Mastrorillo}},
  \bibinfo{journal}{PLoS ONE} \textbf{\bibinfo{volume}{9}},
  \bibinfo{pages}{e97331} (\bibinfo{year}{2014}).

\bibitem[{\citenamefont{Kivel{\"a} et~al.}(2014)\citenamefont{Kivel{\"a},
  Arenas, Barthelemy, Gleeson, Moreno, and Porter}}]{kivela2014multilayer}
\bibinfo{author}{\bibfnamefont{M.}~\bibnamefont{Kivel{\"a}}},
  \bibinfo{author}{\bibfnamefont{A.}~\bibnamefont{Arenas}},
  \bibinfo{author}{\bibfnamefont{M.}~\bibnamefont{Barthelemy}},
  \bibinfo{author}{\bibfnamefont{J.~P.} \bibnamefont{Gleeson}},
  \bibinfo{author}{\bibfnamefont{Y.}~\bibnamefont{Moreno}}, \bibnamefont{and}
  \bibinfo{author}{\bibfnamefont{M.~A.} \bibnamefont{Porter}},
  \bibinfo{journal}{Journal of Complex Networks} \textbf{\bibinfo{volume}{2}},
  \bibinfo{pages}{203} (\bibinfo{year}{2014}).

\bibitem[{\citenamefont{Mount}(2010)}]{flickr1}
\bibinfo{author}{\bibfnamefont{B.}~\bibnamefont{Mount}} (\bibinfo{year}{2010}),
  \bibinfo{note}{available from: \url{http://sfgeo.org/data/tourist-local}}.

\bibitem[{\citenamefont{Thomee et~al.}(2015)\citenamefont{Thomee, Shamma,
  Friedland, Elizalde, Ni, Poland, Borth, and Li}}]{thomee2015yfcc100m}
\bibinfo{author}{\bibfnamefont{B.}~\bibnamefont{Thomee}},
  \bibinfo{author}{\bibfnamefont{D.~A.} \bibnamefont{Shamma}},
  \bibinfo{author}{\bibfnamefont{G.}~\bibnamefont{Friedland}},
  \bibinfo{author}{\bibfnamefont{B.}~\bibnamefont{Elizalde}},
  \bibinfo{author}{\bibfnamefont{K.}~\bibnamefont{Ni}},
  \bibinfo{author}{\bibfnamefont{D.}~\bibnamefont{Poland}},
  \bibinfo{author}{\bibfnamefont{D.}~\bibnamefont{Borth}}, \bibnamefont{and}
  \bibinfo{author}{\bibfnamefont{L.-J.} \bibnamefont{Li}},
  \bibinfo{journal}{arXiv preprint arXiv:1503.01817}  (\bibinfo{year}{2015}).

\bibitem[{\citenamefont{Twitter}(2013)}]{twitterapi}
\bibinfo{author}{\bibnamefont{Twitter}} (\bibinfo{year}{2013}),
  \bibinfo{note}{available from:
  \url{https://dev.twitter.com/streaming/public}}.

\bibitem[{\citenamefont{Bojic et~al.}(2015)\citenamefont{Bojic, Massaro, Belyi,
  Sobolevsky, and Ratti}}]{bojic2015choosing}
\bibinfo{author}{\bibfnamefont{I.}~\bibnamefont{Bojic}},
  \bibinfo{author}{\bibfnamefont{E.}~\bibnamefont{Massaro}},
  \bibinfo{author}{\bibfnamefont{A.}~\bibnamefont{Belyi}},
  \bibinfo{author}{\bibfnamefont{S.}~\bibnamefont{Sobolevsky}},
  \bibnamefont{and} \bibinfo{author}{\bibfnamefont{C.}~\bibnamefont{Ratti}}, in
  \emph{\bibinfo{booktitle}{Social Informatics}}, edited by
  \bibinfo{editor}{\bibfnamefont{T.-Y.} \bibnamefont{Liu}},
  \bibinfo{editor}{\bibfnamefont{C.~N.} \bibnamefont{Scollon}},
  \bibnamefont{and} \bibinfo{editor}{\bibfnamefont{W.}~\bibnamefont{Zhu}}
  (\bibinfo{publisher}{Springer International Publishing},
  \bibinfo{year}{2015}), vol. \bibinfo{volume}{9471} of
  \emph{\bibinfo{series}{Lecture Notes in Computer Science}}, pp.
  \bibinfo{pages}{194--208}, ISBN \bibinfo{isbn}{978-3-319-27432-4},
  \urlprefix\url{http://dx.doi.org/10.1007/978-3-319-27433-1_14}.

\bibitem[{\citenamefont{{United Nations}}(2015)}]{unMigration}
\bibinfo{author}{\bibnamefont{{United Nations}}} (\bibinfo{year}{2015}),
  \bibinfo{note}{available from:
  \url{http://www.un.org/en/development/desa/population/migration/data/index.shtml}}.

\bibitem[{\citenamefont{Hensel}(2009)}]{hensel2009icow}
\bibinfo{author}{\bibfnamefont{P.~R.} \bibnamefont{Hensel}}
  (\bibinfo{year}{2009}), \bibinfo{note}{available from:
  \url{http://www.paulhensel.org/icowcol.html}}.

\bibitem[{\citenamefont{Infoplease}(2015)}]{Languages}
\bibinfo{author}{\bibnamefont{Infoplease}} (\bibinfo{year}{2015}),
  \bibinfo{note}{available from:
  \url{http://www.infoplease.com/ipa/A0855611.html}}.

\bibitem[{\citenamefont{{United Nations}}(2014)}]{UNCOMTRADE}
\bibinfo{author}{\bibnamefont{{United Nations}}} (\bibinfo{year}{2014}),
  \bibinfo{note}{available from: \url{http://comtrade.un.org}}.

\bibitem[{\citenamefont{Zipf}(1946)}]{Gravity1946Zipf}
\bibinfo{author}{\bibfnamefont{G.~K.} \bibnamefont{Zipf}},
  \bibinfo{journal}{American sociological review} pp. \bibinfo{pages}{677--686}
  (\bibinfo{year}{1946}).

\bibitem[{\citenamefont{Barth{\'e}lemy}(2011)}]{Barthelemy2011Spatial}
\bibinfo{author}{\bibfnamefont{M.}~\bibnamefont{Barth{\'e}lemy}},
  \bibinfo{journal}{Physics Reports} \textbf{\bibinfo{volume}{499}},
  \bibinfo{pages}{1} (\bibinfo{year}{2011}).

\bibitem[{\citenamefont{Simini et~al.}(2012)\citenamefont{Simini, Gonz{\'a}lez,
  Maritan, and Barab{\'a}si}}]{RadiationModel}
\bibinfo{author}{\bibfnamefont{F.}~\bibnamefont{Simini}},
  \bibinfo{author}{\bibfnamefont{M.~C.} \bibnamefont{Gonz{\'a}lez}},
  \bibinfo{author}{\bibfnamefont{A.}~\bibnamefont{Maritan}}, \bibnamefont{and}
  \bibinfo{author}{\bibfnamefont{A.-L.} \bibnamefont{Barab{\'a}si}},
  \bibinfo{journal}{Nature} \textbf{\bibinfo{volume}{484}}, \bibinfo{pages}{96}
  (\bibinfo{year}{2012}).

\bibitem[{\citenamefont{Wilson}(1967)}]{wilson1967statistical}
\bibinfo{author}{\bibfnamefont{A.~G.} \bibnamefont{Wilson}},
  \bibinfo{journal}{Transportation research} \textbf{\bibinfo{volume}{1}},
  \bibinfo{pages}{253} (\bibinfo{year}{1967}).

\bibitem[{\citenamefont{Sagarra et~al.}(2013)\citenamefont{Sagarra, Vicente,
  and D{\"\i}az-Guilera}}]{sagarra2013statistical}
\bibinfo{author}{\bibfnamefont{O.}~\bibnamefont{Sagarra}},
  \bibinfo{author}{\bibfnamefont{C.~P.} \bibnamefont{Vicente}},
  \bibnamefont{and}
  \bibinfo{author}{\bibfnamefont{A.}~\bibnamefont{D{\"\i}az-Guilera}},
  \bibinfo{journal}{Physical Review E} \textbf{\bibinfo{volume}{88}},
  \bibinfo{pages}{062806} (\bibinfo{year}{2013}).

\bibitem[{\citenamefont{Grauwin
  et~al.}(2015{\natexlab{b}})\citenamefont{Grauwin, Szell, Sobolevsky,
  H{\"o}vel, Simini, Vanhoof, Smoreda, Barabasi, and
  Ratti}}]{grauwin2015identifying}
\bibinfo{author}{\bibfnamefont{S.}~\bibnamefont{Grauwin}},
  \bibinfo{author}{\bibfnamefont{M.}~\bibnamefont{Szell}},
  \bibinfo{author}{\bibfnamefont{S.}~\bibnamefont{Sobolevsky}},
  \bibinfo{author}{\bibfnamefont{P.}~\bibnamefont{H{\"o}vel}},
  \bibinfo{author}{\bibfnamefont{F.}~\bibnamefont{Simini}},
  \bibinfo{author}{\bibfnamefont{M.}~\bibnamefont{Vanhoof}},
  \bibinfo{author}{\bibfnamefont{Z.}~\bibnamefont{Smoreda}},
  \bibinfo{author}{\bibfnamefont{A.-L.} \bibnamefont{Barabasi}},
  \bibnamefont{and} \bibinfo{author}{\bibfnamefont{C.}~\bibnamefont{Ratti}},
  \bibinfo{journal}{arXiv preprint arXiv:1509.03149}
  (\bibinfo{year}{2015}{\natexlab{b}}).

\bibitem[{\citenamefont{Masucci et~al.}(2013)\citenamefont{Masucci, Serras,
  Johansson, and Batty}}]{masucci2013gravity}
\bibinfo{author}{\bibfnamefont{A.~P.} \bibnamefont{Masucci}},
  \bibinfo{author}{\bibfnamefont{J.}~\bibnamefont{Serras}},
  \bibinfo{author}{\bibfnamefont{A.}~\bibnamefont{Johansson}},
  \bibnamefont{and} \bibinfo{author}{\bibfnamefont{M.}~\bibnamefont{Batty}},
  \bibinfo{journal}{Physical Review E} \textbf{\bibinfo{volume}{88}},
  \bibinfo{pages}{022812} (\bibinfo{year}{2013}).

\bibitem[{\citenamefont{Mucha et~al.}(2010)\citenamefont{Mucha, Richardson,
  Macon, Porter, and Onnela}}]{mucha2010community}
\bibinfo{author}{\bibfnamefont{P.~J.} \bibnamefont{Mucha}},
  \bibinfo{author}{\bibfnamefont{T.}~\bibnamefont{Richardson}},
  \bibinfo{author}{\bibfnamefont{K.}~\bibnamefont{Macon}},
  \bibinfo{author}{\bibfnamefont{M.~A.} \bibnamefont{Porter}},
  \bibnamefont{and} \bibinfo{author}{\bibfnamefont{J.-P.}
  \bibnamefont{Onnela}}, \bibinfo{journal}{Science}
  \textbf{\bibinfo{volume}{328}}, \bibinfo{pages}{876} (\bibinfo{year}{2010}).

\bibitem[{\citenamefont{Tang et~al.}(2012)\citenamefont{Tang, Wang, and
  Liu}}]{tang2012community}
\bibinfo{author}{\bibfnamefont{L.}~\bibnamefont{Tang}},
  \bibinfo{author}{\bibfnamefont{X.}~\bibnamefont{Wang}}, \bibnamefont{and}
  \bibinfo{author}{\bibfnamefont{H.}~\bibnamefont{Liu}}, \bibinfo{journal}{Data
  Mining and Knowledge Discovery} \textbf{\bibinfo{volume}{25}},
  \bibinfo{pages}{1} (\bibinfo{year}{2012}).

\bibitem[{\citenamefont{Newman and Girvan}(2004)}]{newman2004}
\bibinfo{author}{\bibfnamefont{M.}~\bibnamefont{Newman}} \bibnamefont{and}
  \bibinfo{author}{\bibfnamefont{M.}~\bibnamefont{Girvan}},
  \bibinfo{journal}{Physical Review E} \textbf{\bibinfo{volume}{69}},
  \bibinfo{pages}{026113} (\bibinfo{year}{2004}).

\bibitem[{\citenamefont{Newman}(2006)}]{newman2006}
\bibinfo{author}{\bibfnamefont{M.}~\bibnamefont{Newman}},
  \bibinfo{journal}{Proceedings of the National Academy of Sciences}
  \textbf{\bibinfo{volume}{103}}, \bibinfo{pages}{8577} (\bibinfo{year}{2006}).

\bibitem[{\citenamefont{Fortunato and
  Barth{\'e}l{\'e}my}(2007)}]{Fortunato02012007ResolutionLimit}
\bibinfo{author}{\bibfnamefont{S.}~\bibnamefont{Fortunato}} \bibnamefont{and}
  \bibinfo{author}{\bibfnamefont{M.}~\bibnamefont{Barth{\'e}l{\'e}my}},
  \bibinfo{journal}{Proceedings of the National Academy of Sciences}
  \textbf{\bibinfo{volume}{104}}, \bibinfo{pages}{36} (\bibinfo{year}{2007}),
  \eprint{http://www.pnas.org/content/104/1/36.full.pdf+html},
  \urlprefix\url{http://www.pnas.org/content/104/1/36.abstract}.

\bibitem[{\citenamefont{Good et~al.}(2010)\citenamefont{Good, de~Montjoye, and
  Clauset}}]{Good2010PerformanceOfModularity}
\bibinfo{author}{\bibfnamefont{B.~H.} \bibnamefont{Good}},
  \bibinfo{author}{\bibfnamefont{Y.-A.} \bibnamefont{de~Montjoye}},
  \bibnamefont{and} \bibinfo{author}{\bibfnamefont{A.}~\bibnamefont{Clauset}},
  \bibinfo{journal}{Physical Review E} \textbf{\bibinfo{volume}{81}},
  \bibinfo{pages}{046106} (\bibinfo{year}{2010}),
  \urlprefix\url{http://link.aps.org/doi/10.1103/PhysRevE.81.046106}.

\bibitem[{\citenamefont{Arenas et~al.}(2008)\citenamefont{Arenas,
  Fern{\'a}ndez, and G{\'o}mez}}]{Arenas2008Analysis}
\bibinfo{author}{\bibfnamefont{A.}~\bibnamefont{Arenas}},
  \bibinfo{author}{\bibfnamefont{V.}~\bibnamefont{Fern{\'a}ndez}},
  \bibnamefont{and}
  \bibinfo{author}{\bibfnamefont{S.}~\bibnamefont{G{\'o}mez}},
  \bibinfo{journal}{New Journal of Physics} \textbf{\bibinfo{volume}{10}},
  \bibinfo{pages}{053039} (\bibinfo{year}{2008}),
  \urlprefix\url{http://stacks.iop.org/1367-2630/10/i=5/a=053039}.

\bibitem[{\citenamefont{Sobolevsky
  et~al.}(2014{\natexlab{c}})\citenamefont{Sobolevsky, Campari, Belyi, and
  Ratti}}]{Combo}
\bibinfo{author}{\bibfnamefont{S.}~\bibnamefont{Sobolevsky}},
  \bibinfo{author}{\bibfnamefont{R.}~\bibnamefont{Campari}},
  \bibinfo{author}{\bibfnamefont{A.}~\bibnamefont{Belyi}}, \bibnamefont{and}
  \bibinfo{author}{\bibfnamefont{C.}~\bibnamefont{Ratti}},
  \bibinfo{journal}{Physical Review E} \textbf{\bibinfo{volume}{90}},
  \bibinfo{pages}{012811} (\bibinfo{year}{2014}{\natexlab{c}}).

\bibitem[{\citenamefont{Danon et~al.}(2005)\citenamefont{Danon, D\'iaz-Guilera,
  Duch, and Arenas}}]{Danon2005}
\bibinfo{author}{\bibfnamefont{L.}~\bibnamefont{Danon}},
  \bibinfo{author}{\bibfnamefont{A.}~\bibnamefont{D\'iaz-Guilera}},
  \bibinfo{author}{\bibfnamefont{J.}~\bibnamefont{Duch}}, \bibnamefont{and}
  \bibinfo{author}{\bibfnamefont{A.}~\bibnamefont{Arenas}},
  \bibinfo{journal}{Journal of Statistical Mechanics: Theory and Experiment}
  \textbf{\bibinfo{volume}{2005}}, \bibinfo{pages}{P09008}
  (\bibinfo{year}{2005}),
  \urlprefix\url{http://stacks.iop.org/1742-5468/2005/i=09/a=P09008}.

\end{thebibliography}

\end{document}